\documentclass[aip, jcp, reprint, notitlepage, longbibliography, noeprint]{revtex4-2}
\usepackage{amsmath,amsfonts,amssymb,bm} 
\usepackage{xcolor,graphicx,bbold} 
\usepackage{microtype}
\usepackage{epstopdf} 
\usepackage[colorlinks=true,linkcolor=blue,citecolor=blue,urlcolor=blue]{hyperref}
\usepackage[all]{hypcap} 

\newcommand{\Dr}[0]{D_\mathrm{r}}
\newcommand{\Dc}[0]{D_{\rm c}} 
\newcommand{\Deff}[0]{\mathcal{D}}
\newcommand{\Fdc}[0]{F_{\rm dc}}
\newcommand{\Fac}[0]{F_{\rm ac}}
\newcommand{\muc}[0]{\mu_{\rm c}}
\newcommand{\dd}[0]{\mathrm{d}} 
\newcommand{\kB}[0]{k_{\rm B}}
\newcommand{\rvec}[0]{{\bm r}}
\newcommand{\nvec}[0]{{\bm n}}
\newcommand{\Srr}[0]{S_{\bm r \bm r}}
\newcommand{\ua}[0]{u_{\rm a}} 

\begin{document}
\title{Diffusion coefficient and power spectrum of active particles with a microscopically reversible mechanism of self-propelling}

\author{Artem Ryabov} 
\email[]{artem.ryabov@mff.cuni.cz}
\affiliation{Charles University, Faculty of Mathematics and Physics, Department of Macromolecular Physics, V Hole{\v s}ovi{\v c}k{\' a}ch~2, 180~00~Praha~8, Czech Republic} 
\affiliation{Departamento de F\'{\i}sica, Faculdade de Ci\^{e}ncias, Universidade de Lisboa, 
1749-016 Lisboa, Portugal}
\affiliation{Centro de F\'{i}sica Te\'{o}rica e Computacional, 
  Faculdade de Ci\^{e}ncias, Universidade de Lisboa, 1749-016 Lisboa, Portugal}

\author{Mykola Tasinkevych}
\email[]{mykola.tasinkevych@ntu.ac.uk}
\affiliation{SOFT Group, School of Science and Technology, Nottingham Trent University, Clifton Lane, Nottingham NG11 8NS, UK}
\affiliation{Departamento de F\'{\i}sica, Faculdade de Ci\^{e}ncias, Universidade de Lisboa, 
1749-016 Lisboa, Portugal}
\affiliation{Centro de F\'{i}sica Te\'{o}rica e Computacional, 
Faculdade de Ci\^{e}ncias, Universidade de Lisboa, 1749-016 Lisboa, Portugal}

\date{August 05, 2022} 

\begin{abstract}
Catalytically active macromolecules are envisioned as key building blocks in development of artificial nanomotors. However, theory and experiments report conflicting findings regarding their dynamics. The lack of consensus is mostly caused by a limited understanding of specifics of self-propulsion mechanisms at the nanoscale. Here, we study a generic model of a self-propelled nanoparticle that does not rely on a particular mechanism. Instead, its main assumption is the fundamental symmetry of microscopic dynamics of chemical reactions: the principle of microscopic reversibility. Significant consequences of this assumption arise if we subject the particle to the action of an external time-periodic force. The particle diffusion coefficient then becomes enhanced compared to the unbiased dynamics. The enhancement can be controlled by the force amplitude and frequency. We also derive the power spectrum of particle trajectories. Among new effects stemming from the microscopic reversibility are the enhancement of the spectrum at all frequencies and sigmoid-shaped transitions and a peak at characteristic frequencies of rotational diffusion and external forcing. The microscopic reversibility is a generic property of a broad class of chemical reactions, therefore we expect that the presented results will motivate new experimental studies aimed at testing of our predictions. This could provide new insights into dynamics of catalytic macromolecules.
\end{abstract}

\maketitle 

\section{Introduction}
\label{sec:intro}

Active microparticles capable of self-propelled motion are at the forefront of current research in physics, chemistry, and biology.~\cite{Sanchez2015, Bechinger2016, Ramaswamy:JSTAT2017, Zhang/etal:ChemSocRev2017, Palagi/Fischer:NatRevMat2018} The vivid interest in their properties is driven by potential applications, e.g.,  in microscopic robotics,\cite{Palagi/Fischer:NatRevMat2018, Soto:2021, Munos-Landin/etal:SciRobot2021} for targeted transport of drugs\cite{Patra2013, Xu2020, Mitchell/etal:2021} and microcargoes,\cite{Baraban2012}  for cleaning polluted habitats,\cite{Soler2013, Soler2014, Vilela/etal:2022} and working as sensors in biological environments\cite{Wu:2010} to name a few. Inspired by molecular motors, there is an ongoing miniaturization efforts to develop nanosized artificial machines capable to operate, e.g., in the intracellular environment.\cite{Wang:2014} 

However, tracking individual physical processes that can induce the self-propulsion of nanoparticles is a rather challenging task. Instead, several experimental works have focused on average transport characteristics like particle's diffusion constant.\cite{Muddana:2010, Sengupta:2013, Sengupta:2014, Jee/etal:PNAS2018, Ah-Young/etal:2018,Jee:2020,Yuan:2021} Conclusions regarding the nature and existence of self-propulsion are then inferred indirectly based on values of such ensemble-averaged characteristics and confronting them with results for basic theoretical models.\cite{Zhang/Hess:2019} Conversely, theoretical works often conjecture a specific mechanisms of self-propelling and discuss the behaviour of the diffusion constant.\cite{Golestanian:2015,Illien/etal:2017,Agudo-Canalejo:2018a,Agudo-Canalejo/etal:2018} The resulting link between experimental data and theoretical models is thus indirect only. As a consequence, there are often conflicting views\cite{Illien/etal:2017, Kondrat/Popescu:2019, Huan/etal:JPCLett2021, Guenther/etal:JCP2019,Wang/etal:2020, Gunther/etal:2020,Wang/etal:2020:2} on the nature of active motion at these tiny scales and even on whether the catalytic activity of macromolecules can lead to their active self-propelled dynamics.\cite{Guenther/etal:JCP2019,Wang/etal:2020, Gunther/etal:2020}

Despite a number of theoretical and experimental studies of dynamics of active macromolecules,\cite{Feng:2020} the effects of a fundamental premise of nonequilibrium statistical mechanics -- the principle of microscopic reversibility\cite{Tolman:PNAS1925} (MR) remains largely unexplored in this domain. In our recent work,\cite{Ryabov/Tasinkevych:SoftMatt2022} we have analyzed a generic model of the active motion at the nanoscale. The developed model did not rely on a specific self-propulsion mechanisms, but assumed that both the chemical reactions powering the self-propulsion and the translational Brownian motion of the particle comply with the MR principle. We predicted an increased diffusivity and mobility of active nanoparticles compared to a passive particle and to models where MR is not included. As a result of assuming MR, the both parameters become dependent on an amplitude of constant external force applied on the nanoparticle. 

Here, we extend the model of Ref.~\onlinecite{Ryabov/Tasinkevych:SoftMatt2022} by incorporating a time-dependent external force of magnitude $F(t)$ acting upon the active nanoparticle. Motivated by spectroscopic measurements, we assume the force being a superposition of a constant and harmonically oscillating forces, i.e., 
\begin{equation} 
\label{eq:F-amplitude}
F(t) = \Fdc + \Fac \cos(\Omega t+\alpha).
\end{equation}
Our main objective is to describe new qualitative effects stemming from MR in this model. 

First, we will show how parameters of the external force~\eqref{eq:F-amplitude}, i.e., $\Fdc$, $\Fac $ and $\Omega$, can modify the particle's diffusion coefficient. Second, we will derive the power spectrum  of the particle stochastic trajectories. The spectrum  can be measured in single-particle tracking experiments. It contains a more detailed information on the underlying dynamic mechanisms than the diffusion coefficient alone. 

All discussed quantities will be compared with corresponding ones calculated for a reference model, where MR is neglected. Experimental tests of our predictions can decide on the relevance of the MR principle for the self-propulsion at the nanoscale. This in turn, can exclude from consideration (or confirm) a broad range of mechanisms obeying the assumed symmetry, narrowing down possibilities for theoretical modeling. Such tests can also guide an experimental development toward a resolution of existing controversies.

In the following Section~\ref{sec:models}, we introduce our model of a nanoparticle with microscopically reversible propulsion (Sec.~\ref{sec:model-microscopic}), discuss its coarse-grained continuous-space dynamics and estimate all model parameters in accord with relevant experimental data (Sec.~\ref{subsec:parameters}). In addition, we solve the Langevin equations for the time-periodic driving force (Sec.~\ref{subsec:TDconsistent-coarsegrained}), introduce reference models (Sec.~\ref{sec:model-references} and Sec.~\ref{sec:model-references-ABP}), and compare our approach with previous theoretical works (Sec.~\ref{sec:model-review}). In Sec.~\ref{sec:correlations}, the derivation of two-time correlation functions is explained. These are then used to derive and discuss the particle's diffusion coefficient (Sec.~\ref{sec:diffusion-coefficients}) and the power spectrum (Sec.~\ref{sec:spectrum}).

\section{Microscopically reversible active propulsion}
\label{sec:models}

\subsection{Microscopic model}
\label{sec:model-microscopic} 

Consider a particle driven by active propulsion and undergoing  rotational and translational overdamped Brownian motion. Dynamics of the center of mass position $\rvec$ of such a particle is governed by the Langevin equation\cite{Ryabov/Tasinkevych:SoftMatt2022}   
\begin{equation}
\label{eq:Langevin-general}
\frac{\dd \rvec }{\dd  t} = \ua (t) \nvec(t) + 
 \mu {\bm F}(t) + \sqrt{2 D }\, {\bm \xi}(t). 
\end{equation} 
The last two terms on the right-hand side of~\eqref{eq:Langevin-general} represent overdamped Brownian motion in the external force field $\bm F$, $\mu$ is the mobility, $D=\mu \kB T$ the diffusion coefficient, where $\kB$ is the Boltzmann constant, and $T$ the temperature of ambient environment. Components of zero-mean Gaussian white noise vector ${\bm \xi}(t) = (\xi_x(t), \xi_y(t))$ satisfy  
$\langle \xi_i(t_1) \xi_j(t_2) \rangle = \delta_{ij}\delta(t_1-t_2)$. 

In Eq.~\eqref{eq:Langevin-general},  $\ua(t){\bm n}(t)$ denotes the active propulsion velocity  with magnitude $\ua(t)$ and direction ${\bm n}(t)$ being a unit vector 
\begin{equation}
\label{eq:n}
{\bm n}(t)=( \cos\phi(t) ,\sin\phi(t) ),
\end{equation} 
which we also call as the orientation of the particle. 
Over time, the orientation undergoes rotational diffusion\cite{Han/etal:SCIENCE2006} characterized by the diffusion constant $\Dr$, meaning that the angle $\phi(t)$ itself performs a Brownian motion 
\begin{equation} 
\label{eq:phi-result}
\phi(t) = \phi(0) + \sqrt{2 \Dr } \int_0^t\! \xi_{\rm r}(t')\, \dd t',
\end{equation} 
driven by the delta-correlated zero-mean Gaussian white noise 
$\xi_{\rm r}(t)$. The noises $\xi_{\rm r}(t)$ and ${\bm \xi}(t)$ are statistically independent.

It can be rather challenging to precisely control the particle orientation in experiments with nanoswimmers. Therefore, in  this work, the initial value $\phi(0)$ in~\eqref{eq:phi-result} is assumed to be a random variable homogeneously distributed within the interval $[0,2\pi)$, yielding random initial orientation of the particle.

\begin{figure}[t!]
\centering 
\includegraphics[width=0.68\columnwidth]{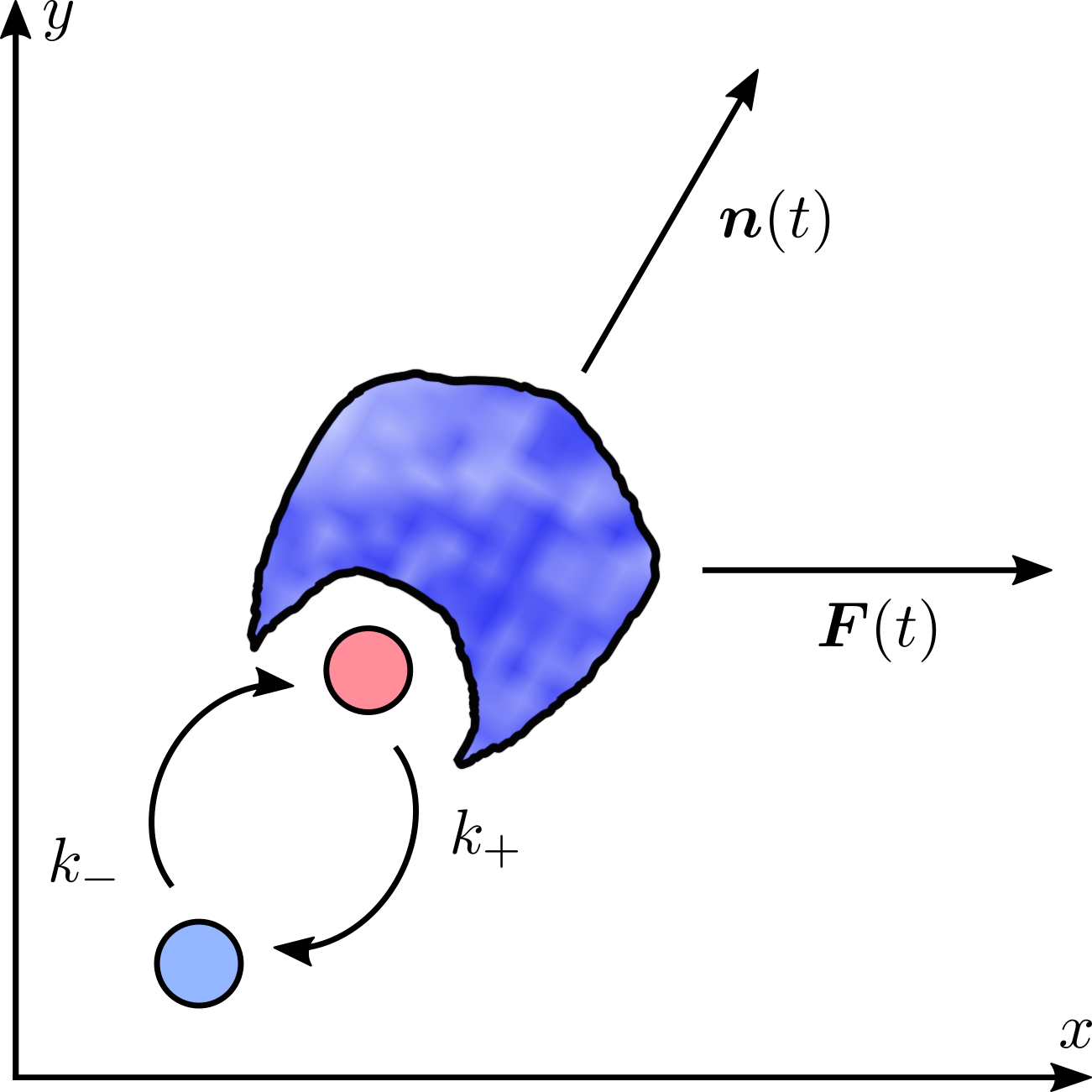}
\caption{Schematics of a chemically active nanoparticle (large blue blob), whose dynamics is governed by the Langevin equation~\eqref{eq:Langevin-general}. A forward chemical reaction, represented as conversion of the small red circle to the light-blue one, occurs with reaction rate $k_+$ and shifts the particle along its instantaneous orientation $\nvec(t)$ by $\delta r$. There can exist a reversed reaction happening with rate $k_-$ and associated with the shift~$(-\delta r)$. Time-reversibility of microscopic dynamics imposes the detailed balance condition~\eqref{eq:detailed-balance} upon the ratio  $k_+/k_-$. In practice, reaction free energies are typically much higher than $\kB T$ ($\Delta G_{\rm r} \gtrsim 5 \kB T$), hence $k_+\gg k_-$ and backward steps cannot be observed. Furthermore, the particle orientation $\nvec(t)$ undergoes rotational diffusion and there is external force $\bm F(t)$ dragging the particle along $x$ direction (horizontal arrow). If such dynamics is observed on coarse-grained macroscopic scales, the force affects magnitude $\ua(t)$ of the active particle velocity according to Eq.~\eqref{eq:ua_macro}. This coupling of chemical and mechanical processes originates from the detailed balance~\eqref{eq:detailed-balance}. It implies remarkable behavior of the diffusion coefficient, Eq.~\eqref{eq:Deff} and Fig.~\ref{fig:Deff}, and spectral power density, Eq.~\eqref{eq:Srr-result}, Figs.~\ref{fig:Srr} and~\ref{fig:Srr-peak}.} 
\label{fig:illustration} 
\end{figure}

In the microscopic model, the magnitude $\ua(t)$ represents stochastic jumps of the particle driven by chemical reactions,\cite{Ryabov/Tasinkevych:SoftMatt2022} see Fig.~\ref{fig:illustration}. We assume that a forward chemical reaction and the corresponding jump from $\rvec$ to $[\rvec+ \nvec(t) \delta r]$ occur with rate $k_+$. The microscopic reversibility\cite{Tolman:PNAS1925, Onsager:1931a, Onsager/Machlup:1953, Astumian:2016} then ensures that there exists also a backward reaction accompanied by the jump from $[\rvec+ \nvec(t) \delta r]$ to $\rvec$ and its rate $k_-$ is related to $k_+$ by the local detailed balance condition\cite{Maes:SciPost2021}  
\begin{equation}
\frac{k_+}{k_-} = 
\exp\!\left( \frac{\Delta G_{\rm r} - \delta W}{\kB T }\right), 
\label{eq:detailed-balance}
\end{equation} 
where  $\Delta  G_{\rm r}$ stands for the reaction free energy and $\delta W$ for the work done on the particle by all external (mechanical, electromagnetic) forces acting on the particle during its jump $\rvec \to [\rvec+ \nvec(t) \delta r]$. 

The condition~\eqref{eq:detailed-balance} follows from reversibility with respect to time-reversal of a more detailed microscopic dynamics (classical Hamiltonian or quantum) upon which our effective stochastic time-evolution~\eqref{eq:Langevin-general} is constructed via elimination of fast degrees of freedom.\cite{Zwanzig:book2001}   

\subsection{Thermodynamically consistent propulsion at macroscale: Estimating parameters of the model}
\label{subsec:parameters}

Miscellaneous single-particle tracking methods,\cite{Metzler/etal:PCCP2014, Jee/etal:PNAS2018, Chen/etal:PNAS2020} pulsed field gradient nuclear magnetic resonance (NMR),\cite{Guenther/etal:JCP2019, Evans:2020, Kaerger/etal:2021, Huan/etal:JPCLett2021} neutron scattering,\cite{Jobic/Theodorou:2007}  and other experimental techniques capable to measure diffusivity of nanoparticles have a common feature: They typically probe the particle dynamics on length scales much larger than $\delta r$ and time-intervals between measurements are longer than the duration of individual reactions. 

When observed on such macroscopic time and length scales, the magnitude of active velocity can be approximated by\cite{Ryabov/Tasinkevych:SoftMatt2022}  
\begin{equation}
\label{eq:ua_macro}
\ua (t) \approx u + \muc F_{n}(t) + \sqrt{2 \Dc}\, \xi_n(t),
\end{equation} 
with  
\begin{equation}
\label{eq:Fn}
F_n(t) = \nvec (t) \cdot {\bm F}(t),  
\end{equation}
being the projection of external force ${\bm F}(t)$ onto the particle orientation $\nvec(t)$. During the coarse-graining procedure, the projection~\eqref{eq:Fn} arises from the work $\delta W \approx F_n(t)\delta r$ present in Eq.~\eqref{eq:detailed-balance}. The constant term $u$ in~\eqref{eq:ua_macro}, on the other hand, is proportional to the reaction free energy per elementary displacement: $(u/\muc) \delta r = \Delta G_{\rm r}$, 
where the mobility  $\muc$ is related to $\Dc$ by the fluctuation-dissipation theorem 
$\muc \kB T = \Dc$
and can be estimated based on the reaction rate constants $k_\pm$ measured in absence of the external force [$\bm F(t)$=0]
\begin{equation} 
\label{eq:Dc-approx}
\Dc \approx \frac{(\delta r)^2}{2} \left( k_+ + k_- \right). 
\end{equation} 
This ``diffusion constant'' sets magnitude of fluctuations in number of chemical reactions powering the particle active dynamics. At the macroscale, these fluctuations are represented by zero-mean Gaussian white noise $\xi_n(t)$ in~\eqref{eq:ua_macro}. 

The derivation of Eq.~\eqref{eq:ua_macro} from the microscopic model can be found in our previous work in Ref.~\onlinecite{Ryabov/Tasinkevych:SoftMatt2022}. 
The essence of this derivation is as follows: In the microscopic model, the active motion is represented by a random walk in the direction of the particle orientation. The forward (backward) steps of length $\delta r$ happen with the rate $k_+$ ($k_-$). The continuum (diffusive) approximation of the dynamics of this random walk process, subjected to the condition that reaction rates $k_\pm$ obey the detailed balance condition~\eqref{eq:detailed-balance}, leads to the Langevin equation with velocity~\eqref{eq:ua_macro}.  

Relation~\eqref{eq:Dc-approx}, $\muc \kB T = \Dc$,  $\mu \kB T = D$, and Stokes' law $\mu=1/6\pi\eta R_{\rm H}$ with $\eta$ being the dynamic viscosity of ambient environment and $R_{\rm H}$ the hydrodynamic radius of nanoparticle, allow us to estimate magnitudes of all the model parameters, which we will use for illustrating our results in Figs.~\ref{fig:Deff},~\ref{fig:Srr}, and partly in Fig.~\ref{fig:Srr-peak}.  
For this purpose, we adopt numerical values inspired by ones reported for catalytically active urease.\cite{Jee/etal:PNAS2018, Ah-Young/etal:2018, Jee:2020} We take  
$k_+ \approx 10^5$~s$^{-1}$,  $\Dr \approx 10^5$~s$^{-1}$, $\delta r \approx 5$~nm, $R_{\rm H}\approx 20$~nm. Furthermore, we neglect $k_-$, set $T \approx 300$~K for the room temperature, and $\eta \approx 8.53\times  10^{-4}$~Ns/m$^2$ for the dynamic viscosity of water at this temperature. As for $u$, surprisingly, its impact on the results turns out to be insignificant even for relatively large velocities like $u=1$~$\mu$m/s.

\subsection{Harmonic driving force}
\label{subsec:TDconsistent-coarsegrained}

In general, the total external driving force ${\bm F}(t)$ in Eq.~\eqref{eq:Fn} can represent any mechanical, electromagnetic, van der Waals, and other forces exerted on the active particle during its dynamics. Here, we assume the $\rvec$-independent force 
${\bm F}(t)= (F(t),0)$ acting along the $x$ coordinate axis with harmonically oscillating amplitude given in Eq.~\eqref{eq:F-amplitude}, 
where $\Omega$ is the angular frequency of oscillations, $\Fdc$ and $\Fac$ are positive constants, and $\alpha$ is the initial phase. To eliminate transient effects caused by a particular value of $\alpha$, we shall average all following results over $\alpha \in [0,2\pi)$.\cite{Ryabov/etal:CNSNS2022} 

We are interested in this particular form of ${\bm F}(t)$ with an outlook on experimental verification of our predictions, since a qualitatively similar external driving appears in many spectroscopic methods.  

For such driving, the projection $F_n(t)$ defined in Eq.~\eqref{eq:Fn} reads 
$F_n(t) = F(t) \cos \phi(t)$. It is multiplied by $\muc \cos\phi(t)$ [$\muc \sin\phi(t)$] in the Langevin equation for $x(t)$ [$y(t)$]. After integrating the Langevin equations subjected to initial conditions $x(0)=y(0)=0$, we get   
\begin{align} 
\label{eq:x-exact}
\begin{split} 
 x(t) & =   
\int_0^t\! \Big\{ [\mu  + \muc \cos^2\! \phi(t')]F(t') + u \cos \phi(t')  \\
&\phantom{=}\quad  +  \sqrt{2 D}\, \xi_{x}(t') +  \cos \phi(t') \sqrt{2 \Dc}\, \xi_{n} (t')  \Big\} \dd t' ,
\end{split} \\ 
\begin{split} 
 y(t) & = \!   
\int_0^t \! \Big[ \muc F(t') \cos \phi(t')\sin \phi(t') + u \sin \phi(t') \\
 & \phantom{=}\quad + \sqrt{2D}\, \xi_{y}(t') + \sin \phi(t') \sqrt{2 \Dc}\, \xi_{n} (t')  \Big] \dd t' . 
\end{split} 
\label{eq:y-exact}
\end{align} 
These integral expressions constitute the starting point for all following derivations. 

\subsection{Reference cases}
\label{sec:model-references}

Assuming the detailed balance condition~\eqref{eq:detailed-balance} has remarkable consequences for the velocity magnitude $\ua(t)$ in Eq.~\eqref{eq:ua_macro}: 
$\ua(t)$ becomes dependent on the external force and the noise strength $\Dc$ in Eq.~\eqref{eq:ua_macro} satisfies the fluctuation-dissipation theorem  $\Dc = \muc \kB T$. 
Setting heuristically 
\begin{equation} 
\label{eq:ua-ABP}
\ua(t) = u = \textrm{const}
\end{equation} 
would eliminate the force-dependence and the noise. Also, for nano-sized particles, this would obliterate the connection between the coarse-grained model based on the Langevin equation~\eqref{eq:Langevin-general} with $\ua(t) = u$ and a microscopic model, where chemical kinetics responsible for the self-propelling is consistent with the principle of microscopic reversibility.\cite{Ryabov/Tasinkevych:SoftMatt2022} 

Formally, neglecting the consequences of MR in our model, in particular the force-dependence in~\eqref{eq:ua_macro}, can be realized by setting $\muc = \Dc /\kB T =0$ in Eq.~\eqref{eq:ua_macro}. Therefore, in the following, we shall refer to the Langevin model of the nanoparticle dynamics with $\ua(t) = u$ as to the one where MR of the active propulsion mechanism is neglected. 

Accordingly, any term occurring in the resulting formulas that would explicitly depend on $\muc$ or $\Dc$, can be regarded as stemming from MR of the active propulsion mechanism. 

The dynamics in the limit $\muc\to 0$ shall serve as a reference case in all discussions of our results.  
Moreover, we will normalize all plotted results by corresponding quantities calculated for $u=0$ and  $\muc = \Dc /\kB T =0$, i.e., by the results for the passive two-dimensional overdamped Brownian motion (BM). 

\subsection{Relation to dynamics of micron-sized particles}
\label{sec:model-references-ABP}

Let us note that the model with dynamics obeying the Langevin equation~\eqref{eq:Langevin-general} with the constant active velocity~\eqref{eq:ua-ABP} is known in the literature simply as the active Brownian particle (ABP) model.~\cite{Erdmann/etal:EPJB2000, Szabo/etal:PRE2006, Peruani/Morelli:PRL2007, Teeffelen/Loewen:PRE2008, tenHagen/etal:2011, Henkes/etal:2011, Bialke/etal:PRL2012, Romanczuk2012, Pototsky/Stark:EPL2012,  Buttinoni/etal:2013, Yang/etal:SOFTMATTER2014, Stenhammar/etal:SOFTMATTER2014, Zottl/Stark:JPCM2016, Das/etal:NJP2018, Malakar/etal:2020, Chaudhuri/Dhar:JSTAT2021} ABP model describes well the dynamics of micron-sized Janus colloidal particles driven by a large number of chemical reactions per second. Due to this large scale, the active motion of micron-sized particles is nearly deterministic and the aforementioned effects of MR are expected to be negligible. That is, for micron-sized particles, the terms in~\eqref{eq:ua_macro} containing  $\muc$ are expected to be much smaller than~$u$.

\subsection{Preliminary work and related models with MR}
\label{sec:model-review}

In Ref.~\onlinecite{Ryabov/Tasinkevych:SoftMatt2022}, we have analyzed the microscopic model of Sec.~\ref{sec:model-microscopic} for the case of a constant (time-independent) external force. In that simpler situation, we could derive exact analytical expressions for first two moments of the particle position. The expressions are valid for the fully microscopic dynamics and all $t$. We have then explained in detail the derivation of the macroscopic limit for active velocity presented in the current Sec.~\ref{subsec:parameters} and found the corresponding moments in this limit. Comparing the results, it turned out that the long-time behavior is qualitatively similar for both levels of description. 

In the current, technically more demanding, time-dependent case, having in mind this equivalency revealed in the previous study,\cite{Ryabov/Tasinkevych:SoftMatt2022} we analyze the macroscopic model only. Another advantage of focusing on the macroscopic limit is that it allows to access a more advanced quantity compared to that discussed in the previous microscopic analysis -- the full power spectrum of an ensemble of stochastic trajectories. 

Let us now sum up some other situations, where certain aspects of MR of the active propulsion were used. In fact, stochastic models of active particles with MR have emerged relatively recently.\cite{Pietzonka/Seifert:2018, Speck:2018} Assuming MR, included into Markovian models by means of the local detailed-balance condition,\cite{Seifert:2011, Speck:2021} is crucial for defining the entropy production of individual stochastic trajectories\cite{Pietzonka/Seifert:2018} within the formalism of stochastic thermodynamics.\cite{Speck:2018} Following works focused on particle's phoretic velocity,\cite{Speck:2019} motility-induced phase separations,\cite{Fisher/etal:2019} and performance of  active heat engines.~\cite{Pietzonka/etal:2019, Speck:2022} 

Another remarkable class of theoretical studies relying on MR focuses on linearized dynamics of both chemical concentrations and mechanical degrees of freedom.\cite{Gaspard/Kapral:2017JCP, gaspar:2018, huang:2018, Gaspard/Kapral:2019a, Gaspard/Kapral:2019, Gaspard/Kapral:2020, DeCorato/Pagonabarraga:2022} MR is then enforced via coupling time-evolutions of the two types of variables by a symmetric matrix of Onsager coefficients, similarly to the formalism of classical linear irreversible thermodynamics.\cite{DeGroot/Mazur:2013} In contrast to these works, in our case, there is no explicit modeling of time-evolution of chemical degrees of freedom. They are treated by means of so called chemiostats, i.e., thermodynamic reservoirs of chemical free energy.

\section{Two-time correlation functions}
\label{sec:correlations} 

In this section, we derive two-time correlation functions, which form mathematical foundations for the physical discussion of diffusion coefficients (Sec.~\ref{sec:diffusion-coefficients}) and the power spectra (Sec.~\ref{sec:spectrum}). 

To proceed with the calculations, we first must evaluate various averages over the particle orientation as given by the angle $\phi(t)$, which performs a Brownian motion, see Eqs.~\eqref{eq:n} and~\eqref{eq:phi-result}. In particular, 
mean values of $\sin  \phi(t)$ and $\cos \phi(t) $ are zero,  
\begin{equation}
\label{eq:sin-cos-mean} 
\langle \sin \phi(t) \rangle =
\langle \cos \phi(t)  \rangle = 0,
\end{equation} 
for all $t\geq 0$ meaning that there is no preferable orientation of the particle, i.e., $\langle \nvec (t) \rangle = {\bm 0}$. 

Furthermore, we have 
\begin{align} 
\label{eq:coscos} 
\begin{split} 
& \langle \cos \phi(t_1) \cos \phi(t_2) \rangle = 
 \langle \sin \phi(t_1) \sin \phi(t_2) \rangle \\ 
& \hspace{8.4em} = \frac{1}{2} e^{- \Dr |t_1-t_2| }, 
\end{split} \\ 
\label{eq:cos2cos2} 
& \langle \cos^2\! \phi(t_1) \cos^2\! \phi(t_2) \rangle = 
\frac{1}{4} + \frac{1}{8} e^{-4 \Dr |t_1-t_2|}, \\
\begin{split} 
& \langle \sin \phi(t_1) \cos \phi(t_1) \sin \phi(t_2) \cos \phi(t_2) \rangle =\frac{1}{8} e^{-4 \Dr |t_1-t_2|} ,  
\end{split}
\end{align} 
and 
\begin{align} 
\label{eq:sincos}
& \langle \sin \phi(t_1) \cos \phi(t_2) \rangle = 0, \\  
\label{eq:cos2cos-sin2cos}
& \langle \cos^2\! \phi(t_1) \cos \phi(t_2) \rangle = \langle \sin^2\! \phi(t_1) \cos \phi(t_2) \rangle = 0, \\
& \langle \sin \phi(t_1) \cos \phi(t_1) \cos \phi(t_2) \rangle =0 ,\\
& \langle \sin \phi(t_1) \cos \phi(t_1) \cos^2\! \phi(t_2) \rangle =0 .
\end{align} 
These identities may be derived by rewriting the goniometric functions in terms of complex exponentials and taking mean values of resulting expressions with respect to the Gaussian distribution of process $\phi(t)$. 

We shall also use the following average over the initial phase of the external force
\begin{equation} 
\label{eq:alfa-average}
\int_0^{2\pi} \!\! \cos(\Omega t_1 + \alpha )\cos(\Omega t_2 + \alpha )\frac{\dd \alpha}{2\pi} = \frac{1}{2} \cos[\Omega (t_1-t_2)],
\end{equation}
and the fact that the mean value of $ \cos(\Omega t + \alpha )$ when $\alpha \in [0,2\pi)$ is zero. 

Averaging Eqs.~\eqref{eq:x-exact} and~\eqref{eq:y-exact} over all noises and $\alpha$, specifically using Eqs.~\eqref{eq:sin-cos-mean}, \eqref{eq:coscos}, and \eqref{eq:sincos}, we arrive at expressions  
\begin{align} 
\label{eq:x-mean}
& \langle x(t) \rangle = \left( \mu + \frac{\muc}{2} \right) \Fdc t , \\ 
& \langle y(t) \rangle = 0, 
\label{eq:y-mean}
\end{align} 
giving mean values of particle coordinates at time~$t$. Thus, the mean position $\langle \rvec (t) \rangle =(\langle x(t) \rangle,\langle y(t) \rangle)$ drifts along the direction of $\bm F(t)$ with the enhanced mobility $(\mu +\muc/2)$ as compared to both the passive Brownian particle and ABP case where $\muc=0$. 

To examine  diffusive dynamics in the presence of such mean drift, it is instructive to focus on the displacement of the particle measured relative to its mean position,  
\begin{equation} 
\Delta \bm \rvec (t) = \bm \rvec (t) - \langle \rvec (t) \rangle,
\end{equation}
with individual coordinates being $\Delta x(t)=x(t)-\langle x(t) \rangle$ and $\Delta y(t)=y(t)-\langle y(t) \rangle = y(t)$. 

In experiments, one most often measures quantities that can be derived from two-time correlation functions 
\begin{subequations} 
\begin{align}
C_{xx}(t_1,t_2) 
= \langle \Delta x(t_1) \Delta x(t_2) \rangle ,\\ 
C_{xy}(t_1,t_2) 
= \langle \Delta x(t_1) \Delta y(t_2) \rangle ,\\ 
C_{yx}(t_1,t_2) 
= \langle \Delta y(t_1) \Delta x(t_2) \rangle ,\\ 
C_{yy}(t_1,t_2) 
= \langle \Delta y(t_1) \Delta y(t_2) \rangle . 
\end{align}
\end{subequations}
Starting with the calculation of $C_{xx}(t_1,t_2)$, 
we subtract $\langle x(t) \rangle$ given in~\eqref{eq:x-mean} from $x(t)$ in Eq.~\eqref{eq:x-exact} and average the product $\Delta x(t_1) \Delta x(t_2)$ over all noises and the initial phase $\alpha$ using Eqs.~\eqref{eq:coscos}, \eqref{eq:cos2cos2}, \eqref{eq:sincos}, \eqref{eq:cos2cos-sin2cos}, and~\eqref{eq:alfa-average}. This yields
\begin{align}
\label{eq:Cxx}
C_{xx}(t_1,t_2) =\; & (2D+\Dc) \min(t_1,t_2) \\ \nonumber 
&\hspace{-5em}
+\!\! \iint\limits_{0\; 0}^{\quad t_1\;t_2}\!\!\! \bigg\{ \frac{u^2}{2} e^{-\Dr |t_1'-t_2'|}
 + \left(\mu + \frac{\muc}{2} \right)^{\!2}\! \frac{\Fac^2}{2} \cos[\Omega(t_1'-t_2')]  \\ \nonumber 
& \hspace{-5em} + \frac{\muc^2}{8}\! \left[\Fdc^2+ \frac{\Fac^2}{2} \cos[\Omega(t_1'-t_2')]\right]\!e^{-4\Dr |t_1'-t_2'|} \bigg\} \dd t_2' \dd t_1'.
\end{align}
Here, terms on the first line result from $\delta$-correlated noises $\xi_x$ and $\xi_n$ in~\eqref{eq:x-exact}, the both terms on the second line would be present also in ABP model (with $\muc=0$), and, the second term also in the harmonically driven passive Brownian motion ($u=0$, $\muc=0$). All the terms on the third line are intrinsic to the present microscopically reversible model and would occur neither in ABP nor in BM models.   

Calculation of the two-time correlation function $C_{yy}(t_1,t_2)$ proceeds along similar lines. The result reads 
\begin{align} 
\label{eq:Cyy}
C_{yy}(t_1,t_2) =\; &  C_{xx}(t_1,t_2) \\ \nonumber  
&-\!\! \iint\limits_{0\; 0}^{\quad t_1\;t_2}\!\!\! 
 \left(\mu + \frac{\muc}{2} \right)^{\!2}\! \frac{\Fac^2}{2} \cos[\Omega(t_1'-t_2')]  \dd t_2' \dd t_1'.
\end{align}
That is, $C_{yy}$ differs from $C_{xx}$ merely by the second term on the second line of Eq.~\eqref{eq:Cxx}. 

For the sake of further analysis, we have solved all integrals in Eqs.~\eqref{eq:Cxx} and~\eqref{eq:Cyy}. The resulting (somewhat extensive) expressions are presented in Appendix, see Eqs.~\eqref{eq:Cxx-explicit} and~\eqref{eq:Cyy-explicit}. 

Finally, after averaging the products $\Delta x(t_1) \Delta y(t_2)$ and $\Delta y(t_1) \Delta x(t_2)$, for the cross-correlations, we get 
\begin{equation}
\label{eq:CxyCyx0}
C_{xy}(t_1,t_2) =C_{yx}(t_1,t_2) = 0.
\end{equation}

\section{Diffusion coefficients}
\label{sec:diffusion-coefficients} 

Sum of results~\eqref{eq:Cxx} and \eqref{eq:Cyy} for $C_{xx}$ and $C_{yy}$ evaluated at $t_1=t_2=t$ gives us the mean squared displacement (MSD) of the particle 
\begin{equation}
\label{eq:MSD-CFs}
\langle [\Delta  \rvec (t) ]^2 \rangle  = C_{xx}(t,t)+C_{yy}(t,t). 
\end{equation} 
When studied in experiments, MSD can provide a valuable insight into the type of microscopic propulsion mechanism used by the nanoparticle. Its exact analytical expression contains a superposition of linear, bounded oscillating and exponentially decaying terms that follow from Eqs.~\eqref{eq:Cxx-explicit} and~\eqref{eq:Cyy-explicit} at $t_1=t_2=t$. Let us inspect two experimentally relevant regimes where this result considerably simplifies.  

Rotational diffusion is the fastest diffusive process in our model. It happens on the characteristic time scale $\sim 1/\Dr$.  If the time-resolution in an experiment is high enough to capture the rotational diffusion, then, at short times, one would observe the linear growth of MSD superimposed with the ballistic term,  
\begin{equation}
\label{eq:MSD-short}
\langle [\Delta  \rvec (t) ]^2 \rangle \approx 4\left(D +\frac{\Dc}{2}\right)t + 
\left[u^2 + \frac{\muc^2}{4}\left(\Fdc^2 + \frac{\Fac^2}{2} \right)\right]t^2, 
\end{equation} 
for $t\ll 1/\Dr$. 

Equation~\eqref{eq:MSD-short} gives MSD of very short persistent trajectories. While moving along such a trajectory, the particle essentially does not rotate. Therefore, the diffusive growth of MSD described by the linear term in~\eqref{eq:MSD-short} is caused just by the translational Brownian motion ($D$) and fluctuations in the active propulsion velocity $\ua(t)$ ($\Dc/2$). The magnitude of the ballistic term ($\sim t^2$) is enhanced by force amplitudes $\Fac^2$ and $\Fdc^2$ as compared to the corresponding result for ABP model.\cite{Howse/etal:2007, Dunderdale/etal:2012, Patino/etal:2018} 

On the other hand, experimental techniques such as the pulsed field gradient NMR,\cite{Guenther/etal:JCP2019, Evans:2020, Kaerger/etal:2021, Huan/etal:JPCLett2021} neutron scattering,\cite{Jobic/Theodorou:2007} and various single-particle tracking methods\cite{Metzler/etal:PCCP2014, Jee/etal:PNAS2018, Chen/etal:PNAS2020} can measure the long-time effective diffusion coefficient  
\begin{equation}
\label{eq:Deff-definice}
\Deff = \lim_{t\to\infty} \frac{ \langle [\Delta  \rvec (t) ]^2 \rangle}{4t} . 
\end{equation}
In this limit, we have  
\begin{equation}
\label{eq:MSD-isotropy}
\lim_{t\to\infty} \frac{ \langle [\Delta x(t) ]^2 \rangle}{t}
=\lim_{t\to\infty} \frac{ \langle [\Delta y(t) ]^2 \rangle}{t} 
=\lim_{t\to\infty} \frac{ \langle [\Delta \rvec(t) ]^2 \rangle}{2t}, 
\end{equation}
i.e., the diffusive spreading of the particle probability density function around its mean position is isotropic. This holds despite the fact that the driving force $\bm F(t)$ breaks spatial isotropy of the problem by dragging the particle along the $x$ axis. 

By evaluating the limit in~\eqref{eq:Deff-definice} we get the effective diffusion coefficient
\begin{align}
\label{eq:Deff}
\begin{split}
\Deff =\ &  D +\frac{\Dc}{2} 
+ \frac{u^2}{2\Dr} + \frac{(\muc \Fdc)^2}{32 \Dr} \\ 
& + \frac{(\muc \Fac)^2}{4\Dr} \frac{\Dr^2}{(4 \Dr)^2+\Omega^2}. 
\end{split}
\end{align} 
The long-time diffusion coefficient $\Deff$ depends on the constant part of active velocity $u$, on the magnitude and frequency of the external force $\bm F(t)$, and on the rotational diffusion constant $\Dr$ in a rather nontrivial manner. Physical origins of individual contributions to $\Deff$ are as follows. 

The contribution $(D+\Dc/2)$ is identical to the short-time diffusivity in Eq.~\eqref{eq:MSD-short}. Its first part reflects the passive translational Brownian motion, for which we have  
\begin{equation}
\label{eq:Deff-BM}
\Deff_{\rm \scriptscriptstyle BM} =  D 
\end{equation}
even in the presence of the external harmonic driving. The second part, $\Dc/2$, arises due to fluctuations in magnitude of the active velocity $\ua (t)$, Eq.~\eqref{eq:ua_macro}. 

The $u$-dependent term in~\eqref{eq:Deff} quantifies uncertainty of the particle position stemming from the rotational diffusion of the constant part $u$ of the active velocity $\ua(t)$. This contribution vanishes with increasing $\Dr$ (at fixed $u$) since the rapidly rotating direction of the velocity has hardly any effect on the translational motion. This term appears also in the effective diffusion coefficient for the harmonically driven ABP model: 
\begin{equation} 
\label{eq:Deff-ABP}
\Deff_{\rm \scriptscriptstyle ABP} =  D + \frac{u^2}{2\Dr} . 
\end{equation}
Remarkably, the value of $u^2/2\Dr$ is rather low as compared to $D$ of a nanoparticle. Using parameter estimates from Sec.~\ref{subsec:parameters}, we get $D\approx 1.3\times 10^{-11}$~m/s$^2$ and $u^2/2\Dr \approx  5\times 10^{-18}$~m/s$^2$ at $u=1$~$\mu$m/s, which is a relatively large value of $u$ when related to particle's hydrodynamic radius $R_{\rm H}\approx 20$~nm. The both terms become comparable at $u\approx 1.6$~mm/s that would correspond to a nanoparticle traveling at staggering $80000\, R_{\rm H}$ per second. The fundamental reason for $u^2/2\Dr$ being vanishingly small as compared to $D$ is fast rotational diffusion of small particles since the rotational diffusion coefficient scales as $\Dr \sim 1/R_{\rm H}^3$. 

Although the $\Fdc$-dependent term in~\eqref{eq:Deff} might seem similar to the $u$-dependent one [both $(\muc \Fdc)$ and $u$ have dimensions of velocity], there is a significant difference in physical origins between the two. While the velocity $u\nvec(t)$ rotates with the particle, the line of action of the external force $\bm F(t)$ is fixed in space  (parallel with the $x$ axis). The force, however, influences the magnitude of active velocity $\ua(t)$, as described by the projection $F_n(\rvec,t)$ in Eq.~\eqref{eq:ua_macro}. When $\nvec(t)$ and $\bm F(t)$ are parallel, the projection attains its maximal magnitude and it vanishes for perpendicular orientation of the two vectors. Because  $\nvec(t)$ rotates erratically, the stochastic changes of $F_n(\rvec,t)$ enhance uncertainty of the particle position as quantified by~$\Deff$. 

\begin{figure}[t!]
\centering 
\includegraphics[scale=1]{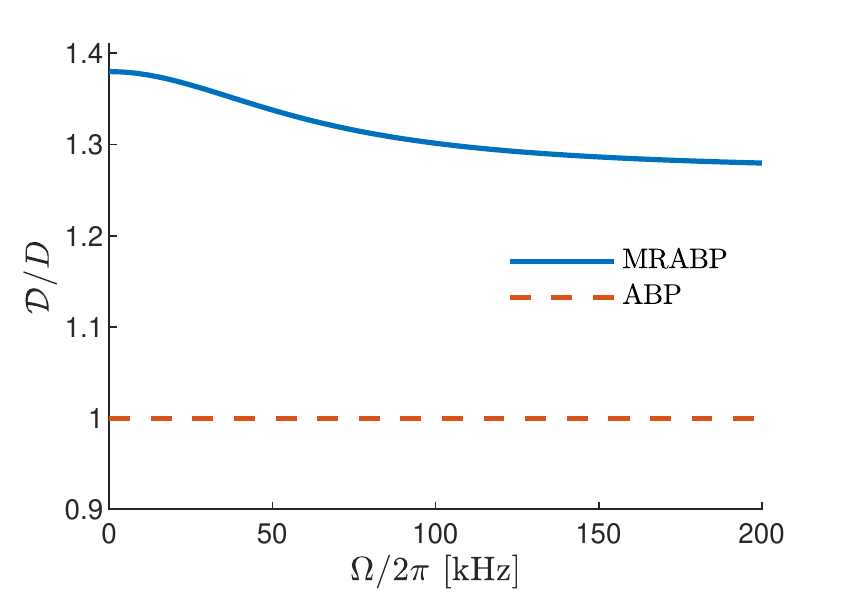}
\caption{Diffusion coefficients of active nanoparticles driven by an external time-periodic force with the amplitude~\eqref{eq:F-amplitude} plotted as functions of driving frequency $\Omega$. The solid line represents $\Deff$ from Eq.~\eqref{eq:Deff} for the model with microscopically reversible active propulsion (MRABP) illustrated in Fig.~\ref{fig:illustration}. The dashed line marks the corresponding result for ABP model where the microscopic reversibility is neglected, cf.\ $\Deff_{\rm \scriptscriptstyle ABP}$ in Eq.~\eqref{eq:Deff-ABP}. Plotted values are scaled by the diffusion constant $D$ of a passive Brownian particle. The reason for force-dependent enhancement of $\Deff$ compared to $D$ and $\Deff_{\rm \scriptscriptstyle ABP}$ can be traced back to the detailed balance condition~\eqref{eq:detailed-balance}. At the nanoscale, $\Deff_{\rm \scriptscriptstyle ABP}$ and $D$ nearly coincide as discussed after Eq.~\eqref{eq:Deff-ABP}. Parameters used are those estimated in Sec.~\ref{subsec:parameters} and we set $\Fdc = \Fac=10$~pN.} 
\label{fig:Deff}
\end{figure}

The last, $\Fac$-dependent, term in~\eqref{eq:Deff} appears for analogous physical reasons. Notably, its magnitude can be controlled by the frequency $\Omega$ of the external force. This dependence is reflected in Fig.~\ref{fig:Deff} showing $\Deff$ from Eq.~\eqref{eq:Deff} (solid line) and $\Deff_{\rm \scriptscriptstyle ABP}$ from~\eqref{eq:Deff-ABP} (dashed), both being normalized by the diffusion coefficient of the overdamped Brownian motion~\eqref{eq:Deff-BM}.

Interestingly, the force-dependent contributions to $\Deff$, whose fundamental origins can be traced back to the detailed balance condition~\eqref{eq:detailed-balance}, provide a way to control the diffusivity by applying external forcing upon the particle. This effect is a direct consequence of the microscopic reversibility of active propulsion mechanism. Therefore, it is missing in the ABP model and in the passive Brownian motion. 

Testing such dependence in an experiment may provide a hint regarding microscopic reversibility of the underlying active propulsion mechanism. The zero-mean harmonic force seems to be well-suited for such purposes because it does not induce a net mean displacement of the particle.

\section{Power spectrum}
\label{sec:spectrum} 

The power spectrum of an ensemble of long trajectories,\cite{Krapf/etal:NJP2018, Squarcini/etal:NJP2022}
\begin{equation}
\Srr(\omega)  = \lim_{\tau\to\infty} \frac{1}{\tau} 
\left\langle \left| 
\int_0^\tau\! \Delta \rvec(t) e^{i\omega t} \dd t 
\right|^2 \right\rangle , 
\end{equation}
bares a more-detailed information on the particle dynamics than MSD~\eqref{eq:MSD-CFs} does. 
Moreover, the power spectrum frequently attains a simple form, where individual terms characterize various diffusive mechanisms involved. 

We will derive $S_{\rvec \rvec}(\omega)$ by breaking it down into two parts:     
$\Srr(\omega)=S_{xx}(\omega)+S_{yy}(\omega)$, with the marginal power spectrum of $\Delta x(t)$ given by  
\begin{align}
\label{eq:Sxx-def}
\begin{split}
S_{xx}(\omega) & = \lim_{\tau\to\infty} \frac{1}{\tau} 
\left\langle \left| 
\int_0^\tau\! \Delta x(t) e^{i\omega t} \dd t 
\right|^2 \right\rangle \\  
& =\lim_{\tau\to\infty} \frac{1}{\tau} \iint\limits_{0\; 0}^{\quad \tau\ \tau} C_{xx}(t_1,t_2)e^{i\omega(t_1-t_2)} \dd t_1 \dd t_2,  
\end{split}
\end{align}
and $S_{yy}(\omega)$ being defined similarly for $\Delta y(t)$.
To evaluate $S_{xx}(\omega)$, we perform the double integration in the second line of Eq.~\eqref{eq:Sxx-def} inserting there the exact expression~\eqref{eq:Cxx-explicit} for $C_{xx}(t_1,t_2)$. Carrying out a similar calculation for $S_{yy}(\omega)$, we find the isotropy relation 
\begin{equation}
\label{eq:Srr-isotropy}
S_{xx}(\omega)=S_{yy}(\omega) ,
\end{equation} 
analogous to Eqs.~\eqref{eq:MSD-isotropy} for MSDs in the long-time limit. 
Eventually, we arrive at the final result 
\begin{widetext}
\begin{align}
\label{eq:Srr-result}
\begin{split}
\Srr(\omega) =\ & \left( D +\frac{\Dc}{2} \right) \frac{8}{\omega^2} 
+ \frac{u^2}{\Dr}\left( \frac{4}{\omega^2} -   \frac{2}{\Dr^2+\omega^2} \right)
+ \frac{(\muc \Fdc)^2}{16 \Dr}\left( \frac{4}{\omega^2}  
- \frac{2}{(4\Dr)^2+\omega^2}\right) \\
&+(\muc \Fac)^2 \frac{\Dr }{(4 \Dr)^2+\Omega^2}\left( \frac{2}{\omega^2} -
\frac{(4\Dr)^2+\omega^2-3\Omega^2}{(4\Dr)^4 +(\omega^2-\Omega^2)^2 + 2(4\Dr)^2 (\Omega^2+\omega^2)}\right).
\end{split} 
\end{align}
\end{widetext} 

All dynamic processes contributing to particle's diffusivity are reflected in $\Deff$ in Eq.~\eqref{eq:Deff}. In addition to their magnitudes that enter the expression for $\Deff$, the power spectrum~\eqref{eq:Srr-result}  resolves corresponding characteristic time scales on which the individual processes happen. 

The first, $(D+\Dc/2)$-dependent part of the result~\eqref{eq:Srr-result} has identical functional form with the power spectrum of two-dimensional overdamped Brownian motion: 
\begin{equation}
\label{eq:Srr-BM}
\Srr^{\rm ({\scriptscriptstyle BM})}(\omega)=\frac{8D}{\omega^2}.    
\end{equation}
The power-law dependence on $\omega$ with exponent $2$ reflects the typical Brownian scaling of position with time. 
At low frequencies ($\omega\to 0$), corresponding to $t\to\infty$ limit in the time domain, the whole expression~\eqref{eq:Srr-result} reduces to just such a power law,
\begin{equation} 
\label{eq:Srr-w0}
\Srr(\omega) \approx \frac{8\Deff}{\omega^2},
\qquad \omega\to 0,
\end{equation} 
where the effective diffusion coefficient $\Deff$ is that from Eq.~\eqref{eq:Deff}. 

The $u$-dependent term in~\eqref{eq:Srr-result} consists of a combination of the Brownian power-law dependence on $\omega$ and the Lorentzian function describing the rotational diffusion of the particle orientation. As the damping rate in the Lorentzian part of this term, there is the characteristic ``frequency'' (inverse time scale) $\Dr$ for this process to happen.

\begin{figure}[t!]
\centering 
\includegraphics[scale=1]{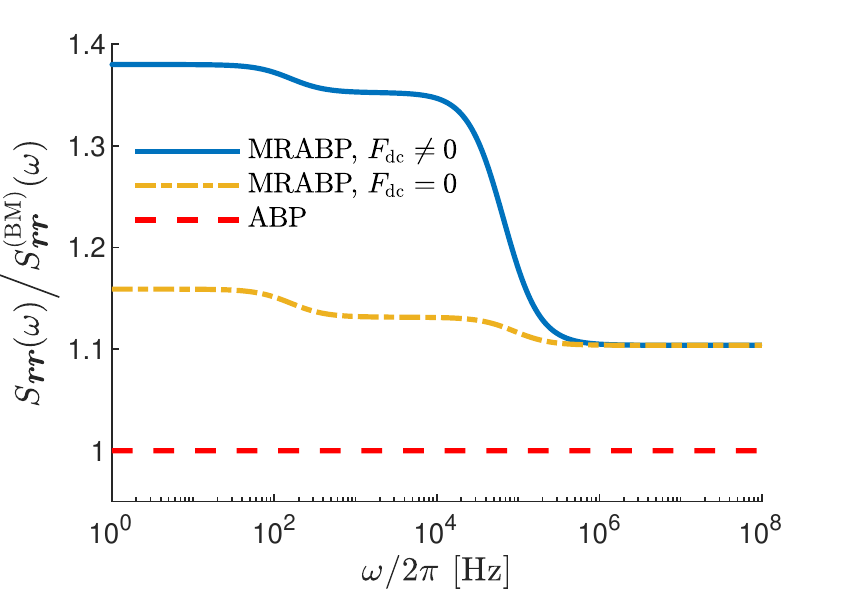}
\caption{Spectral power densities of active nanoparticles driven by the external time-periodic force with amplitude~\eqref{eq:F-amplitude}. Plotted spectra are normalized by $\Srr^{\rm ({\scriptscriptstyle BM})}(\omega)$, the spectrum of overdamped Brownian motion given in Eq.~\eqref{eq:Srr-BM}. Parameters used are those estimated in Sec.~\ref{subsec:parameters}, $\Fac = 10$~pN, and $\Omega=10^3$~s$^{-1}$. For the active particle with microscopically reversible propulsion (MRABP), we plot $\Srr(\omega)$ from Eq.~\eqref{eq:Srr-result} for $\Fdc = 10$~pN (solid line) and $\Fdc = 0$ (dashed-dotted line). The nearly constant dashed line represents the power spectrum $\Srr^{\rm ({\scriptscriptstyle ABP})}(\omega)$ for ABP model where the microscopic reversibility of active propulsion is neglected, see Eq.~\eqref{eq:Srr-ABP}. The $\Fac$-dependent term in~\eqref{eq:Srr-result} causes the sigmoid-shaped changes of MRABP curves near $\omega \approx \Omega$ and $\omega \approx \Dr$ observed at $\Fdc=0$. The $\Fdc$-dependent term enhances the variation of the solid line near $\omega \approx \Dr$ as compared to the $\Fdc=0$ case.} 
\label{fig:Srr}
\end{figure} 

Such $u$-dependent term is also the only additional one that appears in the power spectrum of a harmonically driven ABP model:
\begin{equation}
\label{eq:Srr-ABP}
\Srr^{({\rm {\scriptscriptstyle ABP}})}(\omega) =  \frac{8 D}{\omega^2} 
+ \frac{u^2}{\Dr}\left( \frac{4}{\omega^2} -   \frac{2}{\Dr^2+\omega^2} \right).
\end{equation}
At the nanoscale, its magnitude is rather negligible when compared to the Brownian part $8D/\omega^2$, i.e., $\Srr^{({\rm {\scriptscriptstyle ABP}})}(\omega) \approx \Srr^{({\rm {\scriptscriptstyle BM}})}(\omega)$ holds. The reason for this is the large $\Dr$ of small particles as discussed in details in the paragraph after Eq.~\eqref{eq:Deff-ABP}. 

The $\Fdc$-dependent contribution in~\eqref{eq:Srr-result} has an analogous form as the $u$-dependent one. Yet, the characteristic time scale ($1/4\Dr$) for the process it represents is 4 times shorter than that of the rotational diffusion ($1/\Dr$) occurring in the previous term. The time $1/4\Dr$ can be roughly understood as the decay time of auto-correlations of the product $\nvec(t)F_n(\rvec ,t)$ in Eq.~\eqref{eq:ua_macro} for $\ua(t)$. The individual constituents of this product are already correlated since $F_n(\rvec ,t)$ depends on $\nvec(t)$: As the particle orientation changes due to the rotational diffusion, the magnitude of the force projection onto the instantaneous particle orientation changes as well. Interestingly, this correlation is transferred evenly to power spectra of both coordinates regardless the orientation of the force $\bm F(t)$ in space, viz the isotropy relation~\eqref{eq:Srr-isotropy}. 

Behavior of the $\Fac$-dependent term in~\eqref{eq:Srr-result} with $\omega$ can be rather rich as compared to other terms. Figure~\ref{fig:Srr} illustrates the power spectrum~\eqref{eq:Srr-result} of the microscopically reversible model (MRABP) and the one of a standard periodically driven ABP without the microscopic reversibility of the active propulsion, Eq.~\eqref{eq:Srr-ABP}, for parameters estimated in Sec.~\ref{subsec:parameters} and two values of $\Fdc$. In Fig.~\ref{fig:Srr}, curves marking $\Srr(\omega)$ (solid and dashed-dotted line) indicate that the $\Fac$-dependent term is responsible for two sigmoid-shaped transitions: one close to the driving frequency $\omega \approx \Omega$ and one near $\omega \approx \Dr$. The transitions are visible at for $\Fdc=0$ and $\Fdc=10$~pN. As a result of nonzero $\Fdc$, we observe an enhancement in the low-frequency part of the spectrum. For very high $\omega$, spectra~\eqref{eq:Srr-result} at $\Fdc=0$ and $\Fdc\neq 0$ coincide. Both functions remain significantly larger than $\Srr^{({\rm {\scriptscriptstyle ABP}})}(\omega)$, Eq.~\eqref{eq:Srr-ABP}, which is nearly identical with $\Srr^{({\rm {\scriptscriptstyle BM}})}(\omega)$, Eq.~\eqref{eq:Srr-BM}, for all plotted~$\omega$. 

\begin{figure}[t!]
\centering 
\includegraphics[scale=1]{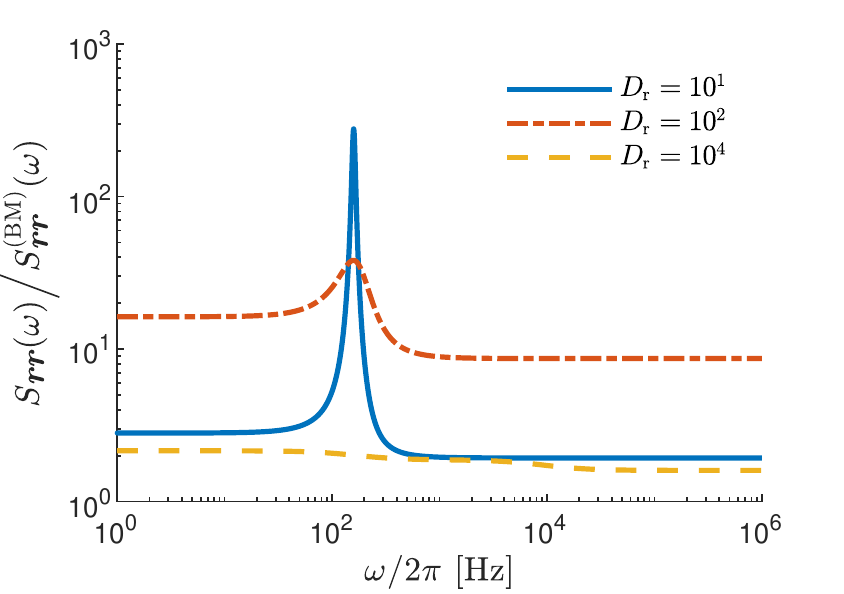}
\caption{Spectral power density~\eqref{eq:Srr-result} of an active particle with microscopically reversible propulsion subjected to the external time-periodic driving. The spectrum is plotted for three different values of  the rotational diffusion coefficient $\Dr$ and normalized by the power spectrum $\Srr^{\rm ({\scriptscriptstyle BM})}(\omega)$ of the overdamped Brownian motion given in Eq.~\eqref{eq:Srr-BM}. Other model parameters are chosen as in Sec.~\ref{subsec:parameters}, $\Fdc=0$, $\Fac=10$~pN, and $\Omega=10^3$~s$^{-1}$. 
The peak in $\Srr(\omega)$ around $\omega\approx \Omega$ develops when $\Dr<\Omega$. Its height can be controlled by $\Dr$, which also changes the background spectral power at other frequencies~$\omega$.} 
\label{fig:Srr-peak}
\end{figure}

Moreover, if parameters $\Fac$, $\Dr$, and $\Omega$ meet certain conditions, the sigmoid-shaped transition around the driving frequency $\omega \approx \Omega$ transforms into a peak, which we demonstrate in Fig.~\ref{fig:Srr-peak}. 
To observe the peak, the $\Fac$-dependent term should attain significant values compared to other terms. This can be achieved either by increasing  $\Fac$ or by a proper choice of $\Dr$, see solid and dashed-dotted curves in Fig.~\ref{fig:Srr-peak} demonstrating the latter option. Also, $\Omega$ should be far enough from low frequencies, where $\Srr(\omega)$ is dominated by the Brownian power-law behavior~\eqref{eq:Srr-w0}. The width and the height of the peak can be controlled by the ``damping'' $4\Dr$. If $4\Dr$ is smaller than $\Omega$, a rather pronounced peak occurs (solid line). The condition $\Omega>\Dr$ means that the external force should oscillate faster than the rotational diffusion happens. Contrary, if the rotational diffusion is faster, the peak vanishes (dashed line in Fig.~\ref{fig:Srr-peak}). 

Let us note that $\Fdc$- and $\Fac$-dependent terms in the power spectrum $\Srr(\omega)$ given in Eq.~\eqref{eq:Srr-result} emerge due to the assumption of the microscopic reversibility of the active propulsion mechanism. Therefore, these terms are naturally missing in the ABP model,  where MR is not taken into account. Accordingly, the derived spectrum $\Srr^{({\rm {\scriptscriptstyle ABP}})}(\omega)$ in Eq.~\eqref{eq:Srr-ABP} for the periodically driven ABP model is identical with the one in case without the external driving force.\cite{Squarcini/etal:NJP2022}

\section{Summary and perspectives}

The principle of microscopic reversibility is inherently related to the time reversal symmetry of microscopic dynamics. It must be enforced whenever the consistency of a studied nonequilibrium stochastic dynamics with the second law of thermodynamics is required. In case of Markov jump processes, the principle is incorporated by means of the local detailed balance condition obeyed by transition rates, while in continuous-space driven diffusions it is implemented via the fluctuation-dissipation relation. Both these ways occur in the modeling of the active propulsion mechanism in the present work. 

As our main results, we have discussed prominent effects stemming from this fundamental principle in the case of a chemically driven active nanoparticle whose propulsion mechanism is made consistent with MR. When an external time-periodic force is acting upon such a particle, its effective diffusion coefficient and the spectral power density are significantly enhanced in comparison to the force-free dynamics and to the corresponding reference model not obeying~MR. 

The diffusion coefficient of such particle contains new contributions that increase its value beyond the one obtained for the reference active Brownian particle model without MR. The magnitude of the enhancement can be externally controlled by varying the amplitude of a constant part of the applied force, the amplitude of a time-periodic part of the force, and the frequency of the force oscillations. 

All the new contributions to the diffusion coefficient have their counterparts in the power spectrum of stochastic trajectories of the nanoparticle. The spectrum also reflects time scales of underlying dynamic processes responsible for these contributions. As new qualitative effects caused by MR in the spectrum, we report sigmoid-shaped transitions and a sharp peak that can occur at the frequency of the external driving and at the characteristic time scale associated with rotational diffusion of the particle. 

Overall, we have selected the external time-periodic driving force having in mind possible spectroscopic verifications of the reported hallmarks of the microscopic reversibility. To this end, we have also chosen values of the model parameters close to the ones of recently studied catalytic nanoparticles. We expect that the results presented here  will motivate new  experimental studies in line with our predictions, which in turn can shed a new light on the nature of chemically driven self-propulsion at the nanoscale. Moreover, our findings can provide new insights into high values of experimentally measured diffusion coefficients of small active particles, view of a potentially significant impact of local (constant and time-dependent) forces on the diffusivity and the power spectrum. 

\section*{Acknowledgements}
We acknowledge financial support from the Portuguese Foundation for Science and Technology (FCT) under Contracts nos. PTDC/FIS-MAC/5689/2020,  UIDB/00618/2020, and UIDP/00618/2020. AR gratefully acknowledges financial support from the Czech Science Foundation (Project No.\ 20-02955J) and from the Department of Physics and Mathematics at Nottingham Trent University (grant no. 01/PHY/-/X1175). Computational resources were supplied by the project ``e-Infrastruktura CZ'' (e-INFRA CZ LM2018140) supported by the Ministry of Education, Youth and Sports of the Czech Republic.

\clearpage 
\appendix*
\section{Exact expressions for correlation functions}
\label{app:correlations}

The analytical expression for $C_{xx}(t_1,t_2)$, which has been given in Eq.~\eqref{eq:Cxx} in terms of double integrals, reads  
\begin{widetext}
\begin{align}
\label{eq:Cxx-explicit}
\begin{split}
C_{xx}(t_1,t_2) =\ & 2 \left[ D +\frac{\Dc}{2} + \frac{u^2}{2\Dr} + \frac{(\muc \Fdc)^2}{32 \Dr} + \frac{(\muc \Fac)^2}{4} \frac{\Dr}{(4 \Dr)^2+\Omega^2} \right] \min(t_1,t_2) \\
& + \frac{u^2}{2\Dr^2}\left( e^{-\Dr t_1}+ e^{-\Dr t_2}- e^{-\Dr|t_2- t_1|}-1 \right)
  + \frac{(\muc \Fdc)^2 }{128\Dr^2} \left( e^{-4\Dr t_1}+ e^{-4\Dr t_2}- e^{-4\Dr|t_2- t_1|}-1 \right)  \\ 
& +\left(\mu + \frac{\muc}{2} \right)^{\! 2}\! \frac{\Fac^2}{2\Omega^2}
\Big\{\!\cos[\Omega (t_2-t_1)]-\cos(\Omega t_1)-\cos(\Omega t_2)+1 \!\Big\}\\ 
& + \frac{(\muc \Fac)^2 }{16} \frac{\Omega^2-(4\Dr)^2}{[(4\Dr)^2+\Omega^2]^2}
\bigg\{ 1- 
e^{-4\Dr|t_2- t_1|} \left[ \cos[\Omega(t_2-t_1)] - \frac{8\Dr \Omega}{(4\Dr)^2-\Omega^2} \sin[\Omega|t_2-t_1|] \right] \\ 
& + e^{-4\Dr t_1} \left[ \cos(\Omega t_1) - \frac{8\Dr \Omega}{(4\Dr)^2-\Omega^2} \sin(\Omega t_1) \right]
 + e^{-4\Dr t_2} \left[ \cos(\Omega t_2) - \frac{8\Dr \Omega}{(4\Dr)^2-\Omega^2} \sin(\Omega t_2) \right] 
\bigg\},
\end{split} 
\end{align}
\end{widetext}
where, on the first line, the expression enclosed in squared brackets  is nothing but the effective diffusion coefficient $\Deff$, Eq.~\eqref{eq:Deff}.
Similarly, we evaluate $C_{yy}(t_1,t_2)$, related to $C_{xx}(t_1,t_2)$ in Eq.~\eqref{eq:Cyy} and get  
\begin{widetext}
\begin{equation}
\label{eq:Cyy-explicit}
C_{yy}(t_1,t_2) = C_{xx}(t_1,t_2) 
- \left(\mu + \frac{\muc}{2} \right)^{\! 2}\! \frac{\Fac^2}{2\Omega^2}
\Big\{\!\cos[\Omega (t_2-t_1)]-\cos(\Omega t_1)-\cos(\Omega t_2)+1 \!\Big\}. 
\end{equation}
\end{widetext}
That is, $C_{yy}(t_1,t_2)$ is given by Eq.~\eqref{eq:Cxx-explicit} after removing all terms displayed on the third line of~\eqref{eq:Cxx-explicit}.  

\clearpage
\section*{References}

\begin{thebibliography}{86}%
\makeatletter
\providecommand \@ifxundefined [1]{%
 \@ifx{#1\undefined}
}%
\providecommand \@ifnum [1]{%
 \ifnum #1\expandafter \@firstoftwo
 \else \expandafter \@secondoftwo
 \fi
}%
\providecommand \@ifx [1]{%
 \ifx #1\expandafter \@firstoftwo
 \else \expandafter \@secondoftwo
 \fi
}%
\providecommand \natexlab [1]{#1}%
\providecommand \enquote  [1]{``#1''}%
\providecommand \bibnamefont  [1]{#1}%
\providecommand \bibfnamefont [1]{#1}%
\providecommand \citenamefont [1]{#1}%
\providecommand \href@noop [0]{\@secondoftwo}%
\providecommand \href [0]{\begingroup \@sanitize@url \@href}%
\providecommand \@href[1]{\@@startlink{#1}\@@href}%
\providecommand \@@href[1]{\endgroup#1\@@endlink}%
\providecommand \@sanitize@url [0]{\catcode `\\12\catcode `\$12\catcode
  `\&12\catcode `\#12\catcode `\^12\catcode `\_12\catcode `\%12\relax}%
\providecommand \@@startlink[1]{}%
\providecommand \@@endlink[0]{}%
\providecommand \url  [0]{\begingroup\@sanitize@url \@url }%
\providecommand \@url [1]{\endgroup\@href {#1}{\urlprefix }}%
\providecommand \urlprefix  [0]{URL }%
\providecommand \Eprint [0]{\href }%
\providecommand \doibase [0]{https://doi.org/}%
\providecommand \selectlanguage [0]{\@gobble}%
\providecommand \bibinfo  [0]{\@secondoftwo}%
\providecommand \bibfield  [0]{\@secondoftwo}%
\providecommand \translation [1]{[#1]}%
\providecommand \BibitemOpen [0]{}%
\providecommand \bibitemStop [0]{}%
\providecommand \bibitemNoStop [0]{.\EOS\space}%
\providecommand \EOS [0]{\spacefactor3000\relax}%
\providecommand \BibitemShut  [1]{\csname bibitem#1\endcsname}%
\let\auto@bib@innerbib\@empty
\bibitem [{\citenamefont {S{\'{a}}nchez}, \citenamefont {Soler},\ and\
  \citenamefont {Katuri}(2015)}]{Sanchez2015}%
  \BibitemOpen
  \bibfield  {author} {\bibinfo {author} {\bibfnamefont {S.}~\bibnamefont
  {S{\'{a}}nchez}}, \bibinfo {author} {\bibfnamefont {L.}~\bibnamefont
  {Soler}},\ and\ \bibinfo {author} {\bibfnamefont {J.}~\bibnamefont
  {Katuri}},\ }\bibfield  {title} {\enquote {\bibinfo {title} {Chemically
  powered micro- and nanomotors},}\ }\href
  {https://doi.org/10.1002/anie.201406096} {\bibfield  {journal} {\bibinfo
  {journal} {Angew. Chem: Int. Ed.}\ }\textbf {\bibinfo {volume} {54}},\
  \bibinfo {pages} {1414} (\bibinfo {year} {2015})}\BibitemShut {NoStop}%
\bibitem [{\citenamefont {Bechinger}\ \emph {et~al.}(2016)\citenamefont
  {Bechinger}, \citenamefont {Di~Leonardo}, \citenamefont {L\"owen},
  \citenamefont {Reichhardt}, \citenamefont {Volpe},\ and\ \citenamefont
  {Volpe}}]{Bechinger2016}%
  \BibitemOpen
  \bibfield  {author} {\bibinfo {author} {\bibfnamefont {C.}~\bibnamefont
  {Bechinger}}, \bibinfo {author} {\bibfnamefont {R.}~\bibnamefont
  {Di~Leonardo}}, \bibinfo {author} {\bibfnamefont {H.}~\bibnamefont
  {L\"owen}}, \bibinfo {author} {\bibfnamefont {C.}~\bibnamefont {Reichhardt}},
  \bibinfo {author} {\bibfnamefont {G.}~\bibnamefont {Volpe}},\ and\ \bibinfo
  {author} {\bibfnamefont {G.}~\bibnamefont {Volpe}},\ }\bibfield  {title}
  {\enquote {\bibinfo {title} {Active particles in complex and crowded
  environments},}\ }\href {https://doi.org/10.1103/RevModPhys.88.045006}
  {\bibfield  {journal} {\bibinfo  {journal} {Rev. Mod. Phys.}\ }\textbf
  {\bibinfo {volume} {88}},\ \bibinfo {pages} {045006} (\bibinfo {year}
  {2016})}\BibitemShut {NoStop}%
\bibitem [{\citenamefont {Ramaswamy}(2017)}]{Ramaswamy:JSTAT2017}%
  \BibitemOpen
  \bibfield  {author} {\bibinfo {author} {\bibfnamefont {S.}~\bibnamefont
  {Ramaswamy}},\ }\bibfield  {title} {\enquote {\bibinfo {title} {Active
  matter},}\ }\href {https://doi.org/10.1088/1742-5468/aa6bc5} {\bibfield
  {journal} {\bibinfo  {journal} {J. Stat. Mech.}\ }\textbf {\bibinfo {volume}
  {2017}},\ \bibinfo {pages} {054002} (\bibinfo {year} {2017})}\BibitemShut
  {NoStop}%
\bibitem [{\citenamefont {Zhang}\ \emph {et~al.}(2017)\citenamefont {Zhang},
  \citenamefont {Luijten}, \citenamefont {Grzybowski},\ and\ \citenamefont
  {Granick}}]{Zhang/etal:ChemSocRev2017}%
  \BibitemOpen
  \bibfield  {author} {\bibinfo {author} {\bibfnamefont {J.}~\bibnamefont
  {Zhang}}, \bibinfo {author} {\bibfnamefont {E.}~\bibnamefont {Luijten}},
  \bibinfo {author} {\bibfnamefont {B.~A.}\ \bibnamefont {Grzybowski}},\ and\
  \bibinfo {author} {\bibfnamefont {S.}~\bibnamefont {Granick}},\ }\bibfield
  {title} {\enquote {\bibinfo {title} {Active colloids with collective mobility
  status and research opportunities},}\ }\href
  {https://doi.org/10.1039/C7CS00461C} {\bibfield  {journal} {\bibinfo
  {journal} {Chem. Soc. Rev.}\ }\textbf {\bibinfo {volume} {46}},\ \bibinfo
  {pages} {5551--5569} (\bibinfo {year} {2017})}\BibitemShut {NoStop}%
\bibitem [{\citenamefont {Palagi}\ and\ \citenamefont
  {Fischer}(2018)}]{Palagi/Fischer:NatRevMat2018}%
  \BibitemOpen
  \bibfield  {author} {\bibinfo {author} {\bibfnamefont {S.}~\bibnamefont
  {Palagi}}\ and\ \bibinfo {author} {\bibfnamefont {P.}~\bibnamefont
  {Fischer}},\ }\bibfield  {title} {\enquote {\bibinfo {title} {Bioinspired
  microrobots},}\ }\href {https://doi.org/10.1038/s41578-018-0016-9} {\bibfield
   {journal} {\bibinfo  {journal} {Nat. Rev. Mater.}\ }\textbf {\bibinfo
  {volume} {3}},\ \bibinfo {pages} {113--124} (\bibinfo {year}
  {2018})}\BibitemShut {NoStop}%
\bibitem [{\citenamefont {Soto}\ \emph {et~al.}(2022)\citenamefont {Soto},
  \citenamefont {Karshalev}, \citenamefont {Zhang}, \citenamefont {Esteban
  Fernandez~de Avila}, \citenamefont {Nourhani},\ and\ \citenamefont
  {Wang}}]{Soto:2021}%
  \BibitemOpen
  \bibfield  {author} {\bibinfo {author} {\bibfnamefont {F.}~\bibnamefont
  {Soto}}, \bibinfo {author} {\bibfnamefont {E.}~\bibnamefont {Karshalev}},
  \bibinfo {author} {\bibfnamefont {F.}~\bibnamefont {Zhang}}, \bibinfo
  {author} {\bibfnamefont {B.}~\bibnamefont {Esteban Fernandez~de Avila}},
  \bibinfo {author} {\bibfnamefont {A.}~\bibnamefont {Nourhani}},\ and\
  \bibinfo {author} {\bibfnamefont {J.}~\bibnamefont {Wang}},\ }\bibfield
  {title} {\enquote {\bibinfo {title} {Smart materials for microrobots},}\
  }\href {https://doi.org/10.1021/acs.chemrev.0c00999} {\bibfield  {journal}
  {\bibinfo  {journal} {Chem. Rev.}\ }\textbf {\bibinfo {volume} {122}},\
  \bibinfo {pages} {5365--5403} (\bibinfo {year} {2022})}\BibitemShut {NoStop}%
\bibitem [{\citenamefont {{n}os Landin}\ \emph {et~al.}(2021)\citenamefont
  {{n}os Landin}, \citenamefont {Fischer}, \citenamefont {Holubec},\ and\
  \citenamefont {Cichos}}]{Munos-Landin/etal:SciRobot2021}%
  \BibitemOpen
  \bibfield  {author} {\bibinfo {author} {\bibfnamefont {S.~M.}\ \bibnamefont
  {{n}os Landin}}, \bibinfo {author} {\bibfnamefont {A.}~\bibnamefont
  {Fischer}}, \bibinfo {author} {\bibfnamefont {V.}~\bibnamefont {Holubec}},\
  and\ \bibinfo {author} {\bibfnamefont {F.}~\bibnamefont {Cichos}},\
  }\bibfield  {title} {\enquote {\bibinfo {title} {Reinforcement learning with
  artificial microswimmers},}\ }\href
  {https://doi.org/10.1126/scirobotics.abd9285} {\bibfield  {journal} {\bibinfo
   {journal} {Sci. Robot.}\ }\textbf {\bibinfo {volume} {6}},\ \bibinfo {pages}
  {eabd9285} (\bibinfo {year} {2021})}\BibitemShut {NoStop}%
\bibitem [{\citenamefont {Patra}\ \emph {et~al.}(2013)\citenamefont {Patra},
  \citenamefont {Sengupta}, \citenamefont {Duan}, \citenamefont {Zhang},
  \citenamefont {Pavlick},\ and\ \citenamefont {Sen}}]{Patra2013}%
  \BibitemOpen
  \bibfield  {author} {\bibinfo {author} {\bibfnamefont {D.}~\bibnamefont
  {Patra}}, \bibinfo {author} {\bibfnamefont {S.}~\bibnamefont {Sengupta}},
  \bibinfo {author} {\bibfnamefont {W.}~\bibnamefont {Duan}}, \bibinfo {author}
  {\bibfnamefont {H.}~\bibnamefont {Zhang}}, \bibinfo {author} {\bibfnamefont
  {R.}~\bibnamefont {Pavlick}},\ and\ \bibinfo {author} {\bibfnamefont
  {A.}~\bibnamefont {Sen}},\ }\bibfield  {title} {\enquote {\bibinfo {title}
  {{Intelligent, self-powered, drug delivery systems}},}\ }\href
  {https://doi.org/10.1039/C2NR32600K} {\bibfield  {journal} {\bibinfo
  {journal} {Nanoscale}\ }\textbf {\bibinfo {volume} {5}},\ \bibinfo {pages}
  {1273} (\bibinfo {year} {2013})}\BibitemShut {NoStop}%
\bibitem [{\citenamefont {Xu}\ \emph {et~al.}(2020)\citenamefont {Xu},
  \citenamefont {Medina-S{\'{a}}nchez}, \citenamefont {Maitz}, \citenamefont
  {Werner},\ and\ \citenamefont {Schmidt}}]{Xu2020}%
  \BibitemOpen
  \bibfield  {author} {\bibinfo {author} {\bibfnamefont {H.}~\bibnamefont
  {Xu}}, \bibinfo {author} {\bibfnamefont {M.}~\bibnamefont
  {Medina-S{\'{a}}nchez}}, \bibinfo {author} {\bibfnamefont {M.~F.}\
  \bibnamefont {Maitz}}, \bibinfo {author} {\bibfnamefont {C.}~\bibnamefont
  {Werner}},\ and\ \bibinfo {author} {\bibfnamefont {O.~G.}\ \bibnamefont
  {Schmidt}},\ }\bibfield  {title} {\enquote {\bibinfo {title} {Sperm
  micromotors for cargo delivery through flowing blood},}\ }\href
  {https://doi.org/10.1021/acsnano.9b07851} {\bibfield  {journal} {\bibinfo
  {journal} {ACS Nano}\ }\textbf {\bibinfo {volume} {14}},\ \bibinfo {pages}
  {2982} (\bibinfo {year} {2020})}\BibitemShut {NoStop}%
\bibitem [{\citenamefont {Mitchell}\ \emph {et~al.}(2021)\citenamefont
  {Mitchell}, \citenamefont {Billingsley}, \citenamefont {Haley}, \citenamefont
  {Wechsler}, \citenamefont {Peppas},\ and\ \citenamefont
  {Langer}}]{Mitchell/etal:2021}%
  \BibitemOpen
  \bibfield  {author} {\bibinfo {author} {\bibfnamefont {M.~J.}\ \bibnamefont
  {Mitchell}}, \bibinfo {author} {\bibfnamefont {M.~M.}\ \bibnamefont
  {Billingsley}}, \bibinfo {author} {\bibfnamefont {R.~M.}\ \bibnamefont
  {Haley}}, \bibinfo {author} {\bibfnamefont {M.~E.}\ \bibnamefont {Wechsler}},
  \bibinfo {author} {\bibfnamefont {N.~A.}\ \bibnamefont {Peppas}},\ and\
  \bibinfo {author} {\bibfnamefont {R.}~\bibnamefont {Langer}},\ }\bibfield
  {title} {\enquote {\bibinfo {title} {Engineering precision nanoparticles for
  drug delivery},}\ }\href {https://doi.org/10.1038/s41573-020-0090-8}
  {\bibfield  {journal} {\bibinfo  {journal} {Nat. Rev. Drug Discov.}\ }\textbf
  {\bibinfo {volume} {20}},\ \bibinfo {pages} {101--124} (\bibinfo {year}
  {2021})}\BibitemShut {NoStop}%
\bibitem [{\citenamefont {Baraban}\ \emph {et~al.}(2012)\citenamefont
  {Baraban}, \citenamefont {Tasinkevych}, \citenamefont {Popescu},
  \citenamefont {Sanchez}, \citenamefont {Dietrich},\ and\ \citenamefont
  {Schmidt}}]{Baraban2012}%
  \BibitemOpen
  \bibfield  {author} {\bibinfo {author} {\bibfnamefont {L.}~\bibnamefont
  {Baraban}}, \bibinfo {author} {\bibfnamefont {M.}~\bibnamefont
  {Tasinkevych}}, \bibinfo {author} {\bibfnamefont {M.~N.}\ \bibnamefont
  {Popescu}}, \bibinfo {author} {\bibfnamefont {S.}~\bibnamefont {Sanchez}},
  \bibinfo {author} {\bibfnamefont {S.}~\bibnamefont {Dietrich}},\ and\
  \bibinfo {author} {\bibfnamefont {O.~G.}\ \bibnamefont {Schmidt}},\
  }\bibfield  {title} {\enquote {\bibinfo {title} {Transport of cargo by
  catalytic {Janus} micro-motors},}\ }\href
  {https://doi.org/10.1039/C1SM06512B} {\bibfield  {journal} {\bibinfo
  {journal} {Soft Matter}\ }\textbf {\bibinfo {volume} {8}},\ \bibinfo {pages}
  {48} (\bibinfo {year} {2012})}\BibitemShut {NoStop}%
\bibitem [{\citenamefont {Soler}\ \emph {et~al.}(2013)\citenamefont {Soler},
  \citenamefont {Magdanz}, \citenamefont {Fomin}, \citenamefont {Sanchez},\
  and\ \citenamefont {Schmidt}}]{Soler2013}%
  \BibitemOpen
  \bibfield  {author} {\bibinfo {author} {\bibfnamefont {L.}~\bibnamefont
  {Soler}}, \bibinfo {author} {\bibfnamefont {V.}~\bibnamefont {Magdanz}},
  \bibinfo {author} {\bibfnamefont {V.~M.}\ \bibnamefont {Fomin}}, \bibinfo
  {author} {\bibfnamefont {S.}~\bibnamefont {Sanchez}},\ and\ \bibinfo {author}
  {\bibfnamefont {O.~G.}\ \bibnamefont {Schmidt}},\ }\bibfield  {title}
  {\enquote {\bibinfo {title} {Self-propelled micromotors for cleaning polluted
  water},}\ }\href {https://doi.org/10.1021/nn405075d} {\bibfield  {journal}
  {\bibinfo  {journal} {ACS Nano}\ }\textbf {\bibinfo {volume} {7}},\ \bibinfo
  {pages} {9611} (\bibinfo {year} {2013})}\BibitemShut {NoStop}%
\bibitem [{\citenamefont {Soler}\ and\ \citenamefont
  {S{\'{a}}nchez}(2014)}]{Soler2014}%
  \BibitemOpen
  \bibfield  {author} {\bibinfo {author} {\bibfnamefont {L.}~\bibnamefont
  {Soler}}\ and\ \bibinfo {author} {\bibfnamefont {S.}~\bibnamefont
  {S{\'{a}}nchez}},\ }\bibfield  {title} {\enquote {\bibinfo {title} {Catalytic
  nanomotors for environmental monitoring and water remediation},}\ }\href
  {https://doi.org/10.1039/C4NR01321B} {\bibfield  {journal} {\bibinfo
  {journal} {Nanoscale}\ }\textbf {\bibinfo {volume} {6}},\ \bibinfo {pages}
  {7175} (\bibinfo {year} {2014})}\BibitemShut {NoStop}%
\bibitem [{\citenamefont {Vilela}\ \emph {et~al.}(2022)\citenamefont {Vilela},
  \citenamefont {Guix}, \citenamefont {Parmar}, \citenamefont {Blanco-Blanes},\
  and\ \citenamefont {S{\' a}nchez}}]{Vilela/etal:2022}%
  \BibitemOpen
  \bibfield  {author} {\bibinfo {author} {\bibfnamefont {D.}~\bibnamefont
  {Vilela}}, \bibinfo {author} {\bibfnamefont {M.}~\bibnamefont {Guix}},
  \bibinfo {author} {\bibfnamefont {J.}~\bibnamefont {Parmar}}, \bibinfo
  {author} {\bibfnamefont {{\'A}.}~\bibnamefont {Blanco-Blanes}},\ and\
  \bibinfo {author} {\bibfnamefont {S.}~\bibnamefont {S{\' a}nchez}},\
  }\bibfield  {title} {\enquote {\bibinfo {title} {Micromotor-in-sponge
  platform for multicycle large-volume degradation of organic pollutants},}\
  }\href {https://doi.org/https://doi.org/10.1002/smll.202107619} {\bibfield
  {journal} {\bibinfo  {journal} {Small}\ }\textbf {\bibinfo {volume} {18}},\
  \bibinfo {pages} {2107619} (\bibinfo {year} {2022})}\BibitemShut {NoStop}%
\bibitem [{\citenamefont {Wu}\ \emph {et~al.}(2010)\citenamefont {Wu},
  \citenamefont {Balasubramanian}, \citenamefont {Kagan}, \citenamefont
  {Manesh}, \citenamefont {Campuzano},\ and\ \citenamefont {Wang}}]{Wu:2010}%
  \BibitemOpen
  \bibfield  {author} {\bibinfo {author} {\bibfnamefont {J.}~\bibnamefont
  {Wu}}, \bibinfo {author} {\bibfnamefont {S.}~\bibnamefont {Balasubramanian}},
  \bibinfo {author} {\bibfnamefont {D.}~\bibnamefont {Kagan}}, \bibinfo
  {author} {\bibfnamefont {K.~M.}\ \bibnamefont {Manesh}}, \bibinfo {author}
  {\bibfnamefont {S.}~\bibnamefont {Campuzano}},\ and\ \bibinfo {author}
  {\bibfnamefont {J.}~\bibnamefont {Wang}},\ }\bibfield  {title} {\enquote
  {\bibinfo {title} {Motion-based {DNA} detection using catalytic
  nanomotors},}\ }\href {https://doi.org/10.1038/ncomms1035} {\bibfield
  {journal} {\bibinfo  {journal} {Nat. Commun.}\ }\textbf {\bibinfo {volume}
  {1}},\ \bibinfo {pages} {36} (\bibinfo {year} {2010})}\BibitemShut {NoStop}%
\bibitem [{\citenamefont {Wang}\ \emph {et~al.}(2014)\citenamefont {Wang},
  \citenamefont {Li}, \citenamefont {Mair}, \citenamefont {Ahmed},
  \citenamefont {Huang},\ and\ \citenamefont {Mallouk}}]{Wang:2014}%
  \BibitemOpen
  \bibfield  {author} {\bibinfo {author} {\bibfnamefont {W.}~\bibnamefont
  {Wang}}, \bibinfo {author} {\bibfnamefont {S.}~\bibnamefont {Li}}, \bibinfo
  {author} {\bibfnamefont {L.}~\bibnamefont {Mair}}, \bibinfo {author}
  {\bibfnamefont {S.}~\bibnamefont {Ahmed}}, \bibinfo {author} {\bibfnamefont
  {T.~J.}\ \bibnamefont {Huang}},\ and\ \bibinfo {author} {\bibfnamefont
  {T.~E.}\ \bibnamefont {Mallouk}},\ }\bibfield  {title} {\enquote {\bibinfo
  {title} {Acoustic propulsion of nanorod motors inside living cells},}\ }\href
  {https://doi.org/https://doi.org/10.1002/anie.201309629} {\bibfield
  {journal} {\bibinfo  {journal} {Angew. Chem., Int. Ed. Engl.}\ }\textbf
  {\bibinfo {volume} {53}},\ \bibinfo {pages} {3201--3204} (\bibinfo {year}
  {2014})}\BibitemShut {NoStop}%
\bibitem [{\citenamefont {Muddana}\ \emph {et~al.}(2010)\citenamefont
  {Muddana}, \citenamefont {Sengupta}, \citenamefont {Mallouk}, \citenamefont
  {Sen},\ and\ \citenamefont {Butler}}]{Muddana:2010}%
  \BibitemOpen
  \bibfield  {author} {\bibinfo {author} {\bibfnamefont {H.~S.}\ \bibnamefont
  {Muddana}}, \bibinfo {author} {\bibfnamefont {S.}~\bibnamefont {Sengupta}},
  \bibinfo {author} {\bibfnamefont {T.~E.}\ \bibnamefont {Mallouk}}, \bibinfo
  {author} {\bibfnamefont {A.}~\bibnamefont {Sen}},\ and\ \bibinfo {author}
  {\bibfnamefont {P.~J.}\ \bibnamefont {Butler}},\ }\bibfield  {title}
  {\enquote {\bibinfo {title} {Substrate catalysis enhances single-enzyme
  diffusion},}\ }\href {https://doi.org/10.1021/ja908773a} {\bibfield
  {journal} {\bibinfo  {journal} {J. Am. Chem. Soc.}\ }\textbf {\bibinfo
  {volume} {132}},\ \bibinfo {pages} {2110--2111} (\bibinfo {year}
  {2010})}\BibitemShut {NoStop}%
\bibitem [{\citenamefont {Sengupta}\ \emph {et~al.}(2013)\citenamefont
  {Sengupta}, \citenamefont {Dey}, \citenamefont {Muddana}, \citenamefont
  {Tabouillot}, \citenamefont {Ibele}, \citenamefont {Butler},\ and\
  \citenamefont {Sen}}]{Sengupta:2013}%
  \BibitemOpen
  \bibfield  {author} {\bibinfo {author} {\bibfnamefont {S.}~\bibnamefont
  {Sengupta}}, \bibinfo {author} {\bibfnamefont {K.~K.}\ \bibnamefont {Dey}},
  \bibinfo {author} {\bibfnamefont {H.~S.}\ \bibnamefont {Muddana}}, \bibinfo
  {author} {\bibfnamefont {T.}~\bibnamefont {Tabouillot}}, \bibinfo {author}
  {\bibfnamefont {M.~E.}\ \bibnamefont {Ibele}}, \bibinfo {author}
  {\bibfnamefont {P.~J.}\ \bibnamefont {Butler}},\ and\ \bibinfo {author}
  {\bibfnamefont {A.}~\bibnamefont {Sen}},\ }\bibfield  {title} {\enquote
  {\bibinfo {title} {Enzyme molecules as nanomotors},}\ }\href
  {https://doi.org/10.1021/ja3091615} {\bibfield  {journal} {\bibinfo
  {journal} {J. Am. Chem. Soc.}\ }\textbf {\bibinfo {volume} {135}},\ \bibinfo
  {pages} {1406--1414} (\bibinfo {year} {2013})}\BibitemShut {NoStop}%
\bibitem [{\citenamefont {Sengupta}\ \emph {et~al.}(2014)\citenamefont
  {Sengupta}, \citenamefont {Spiering}, \citenamefont {Dey}, \citenamefont
  {Duan}, \citenamefont {Patra}, \citenamefont {Butler}, \citenamefont
  {Astumian}, \citenamefont {Benkovic},\ and\ \citenamefont
  {Sen}}]{Sengupta:2014}%
  \BibitemOpen
  \bibfield  {author} {\bibinfo {author} {\bibfnamefont {S.}~\bibnamefont
  {Sengupta}}, \bibinfo {author} {\bibfnamefont {M.~M.}\ \bibnamefont
  {Spiering}}, \bibinfo {author} {\bibfnamefont {K.~K.}\ \bibnamefont {Dey}},
  \bibinfo {author} {\bibfnamefont {W.}~\bibnamefont {Duan}}, \bibinfo {author}
  {\bibfnamefont {D.}~\bibnamefont {Patra}}, \bibinfo {author} {\bibfnamefont
  {P.~J.}\ \bibnamefont {Butler}}, \bibinfo {author} {\bibfnamefont {R.~D.}\
  \bibnamefont {Astumian}}, \bibinfo {author} {\bibfnamefont {S.~J.}\
  \bibnamefont {Benkovic}},\ and\ \bibinfo {author} {\bibfnamefont
  {A.}~\bibnamefont {Sen}},\ }\bibfield  {title} {\enquote {\bibinfo {title}
  {{DNA} polymerase as a molecular motor and pump},}\ }\href
  {https://doi.org/10.1021/nn405963x} {\bibfield  {journal} {\bibinfo
  {journal} {ACS Nano}\ }\textbf {\bibinfo {volume} {8}},\ \bibinfo {pages}
  {2410--2418} (\bibinfo {year} {2014})}\BibitemShut {NoStop}%
\bibitem [{\citenamefont {Jee}\ \emph {et~al.}(2018{\natexlab{a}})\citenamefont
  {Jee}, \citenamefont {Dutta}, \citenamefont {Cho}, \citenamefont {Tlusty},\
  and\ \citenamefont {Granick}}]{Jee/etal:PNAS2018}%
  \BibitemOpen
  \bibfield  {author} {\bibinfo {author} {\bibfnamefont {A.-Y.}\ \bibnamefont
  {Jee}}, \bibinfo {author} {\bibfnamefont {S.}~\bibnamefont {Dutta}}, \bibinfo
  {author} {\bibfnamefont {Y.-K.}\ \bibnamefont {Cho}}, \bibinfo {author}
  {\bibfnamefont {T.}~\bibnamefont {Tlusty}},\ and\ \bibinfo {author}
  {\bibfnamefont {S.}~\bibnamefont {Granick}},\ }\bibfield  {title} {\enquote
  {\bibinfo {title} {Enzyme leaps fuel antichemotaxis},}\ }\href
  {https://doi.org/10.1073/pnas.1717844115} {\bibfield  {journal} {\bibinfo
  {journal} {Proc. Natl. Acad. Sci. USA}\ }\textbf {\bibinfo {volume} {115}},\
  \bibinfo {pages} {14--18} (\bibinfo {year} {2018}{\natexlab{a}})}\BibitemShut
  {NoStop}%
\bibitem [{\citenamefont {Jee}\ \emph {et~al.}(2018{\natexlab{b}})\citenamefont
  {Jee}, \citenamefont {Cho}, \citenamefont {Granick},\ and\ \citenamefont
  {Tlusty}}]{Ah-Young/etal:2018}%
  \BibitemOpen
  \bibfield  {author} {\bibinfo {author} {\bibfnamefont {A.-Y.}\ \bibnamefont
  {Jee}}, \bibinfo {author} {\bibfnamefont {Y.-K.}\ \bibnamefont {Cho}},
  \bibinfo {author} {\bibfnamefont {S.}~\bibnamefont {Granick}},\ and\ \bibinfo
  {author} {\bibfnamefont {T.}~\bibnamefont {Tlusty}},\ }\bibfield  {title}
  {\enquote {\bibinfo {title} {Catalytic enzymes are active matter},}\ }\href
  {https://doi.org/10.1073/pnas.1814180115} {\bibfield  {journal} {\bibinfo
  {journal} {Proc. Natl. Acad. Sci. U.S.A.}\ }\textbf {\bibinfo {volume}
  {115}},\ \bibinfo {pages} {E10812} (\bibinfo {year}
  {2018}{\natexlab{b}})}\BibitemShut {NoStop}%
\bibitem [{\citenamefont {Jee}, \citenamefont {Tlusty},\ and\ \citenamefont
  {Granick}(2020)}]{Jee:2020}%
  \BibitemOpen
  \bibfield  {author} {\bibinfo {author} {\bibfnamefont {A.-Y.}\ \bibnamefont
  {Jee}}, \bibinfo {author} {\bibfnamefont {T.}~\bibnamefont {Tlusty}},\ and\
  \bibinfo {author} {\bibfnamefont {S.}~\bibnamefont {Granick}},\ }\bibfield
  {title} {\enquote {\bibinfo {title} {Master curve of boosted diffusion for 10
  catalytic enzymes},}\ }\href {https://doi.org/10.1073/pnas.2019810117}
  {\bibfield  {journal} {\bibinfo  {journal} {Proc. Natl. Acad. Sci. USA}\
  }\textbf {\bibinfo {volume} {117}},\ \bibinfo {pages} {29435--29441}
  (\bibinfo {year} {2020})}\BibitemShut {NoStop}%
\bibitem [{\citenamefont {Yuan}\ \emph {et~al.}(2021)\citenamefont {Yuan},
  \citenamefont {Liu}, \citenamefont {Wang},\ and\ \citenamefont
  {Ma}}]{Yuan:2021}%
  \BibitemOpen
  \bibfield  {author} {\bibinfo {author} {\bibfnamefont {H.}~\bibnamefont
  {Yuan}}, \bibinfo {author} {\bibfnamefont {X.}~\bibnamefont {Liu}}, \bibinfo
  {author} {\bibfnamefont {L.}~\bibnamefont {Wang}},\ and\ \bibinfo {author}
  {\bibfnamefont {X.}~\bibnamefont {Ma}},\ }\bibfield  {title} {\enquote
  {\bibinfo {title} {Fundamentals and applications of enzyme powered
  micro/nano-motors},}\ }\href
  {https://doi.org/https://doi.org/10.1016/j.bioactmat.2020.11.022} {\bibfield
  {journal} {\bibinfo  {journal} {Bioact. Mater.}\ }\textbf {\bibinfo {volume}
  {6}},\ \bibinfo {pages} {1727--1749} (\bibinfo {year} {2021})}\BibitemShut
  {NoStop}%
\bibitem [{\citenamefont {Zhang}\ and\ \citenamefont
  {Hess}(2019)}]{Zhang/Hess:2019}%
  \BibitemOpen
  \bibfield  {author} {\bibinfo {author} {\bibfnamefont {Y.}~\bibnamefont
  {Zhang}}\ and\ \bibinfo {author} {\bibfnamefont {H.}~\bibnamefont {Hess}},\
  }\bibfield  {title} {\enquote {\bibinfo {title} {Enhanced diffusion of
  catalytically active enzymes},}\ }\href
  {https://doi.org/10.1021/acscentsci.9b00228} {\bibfield  {journal} {\bibinfo
  {journal} {ACS Cent. Sci.}\ }\textbf {\bibinfo {volume} {5}},\ \bibinfo
  {pages} {939--948} (\bibinfo {year} {2019})}\BibitemShut {NoStop}%
\bibitem [{\citenamefont {Golestanian}(2015)}]{Golestanian:2015}%
  \BibitemOpen
  \bibfield  {author} {\bibinfo {author} {\bibfnamefont {R.}~\bibnamefont
  {Golestanian}},\ }\bibfield  {title} {\enquote {\bibinfo {title} {Enhanced
  diffusion of enzymes that catalyze exothermic reactions},}\ }\href
  {https://doi.org/10.1103/PhysRevLett.115.108102} {\bibfield  {journal}
  {\bibinfo  {journal} {Phys. Rev. Lett.}\ }\textbf {\bibinfo {volume} {115}},\
  \bibinfo {pages} {108102} (\bibinfo {year} {2015})}\BibitemShut {NoStop}%
\bibitem [{\citenamefont {Illien}\ \emph {et~al.}(2017)\citenamefont {Illien},
  \citenamefont {Zhao}, \citenamefont {Dey}, \citenamefont {Butler},
  \citenamefont {Sen},\ and\ \citenamefont {Golestanian}}]{Illien/etal:2017}%
  \BibitemOpen
  \bibfield  {author} {\bibinfo {author} {\bibfnamefont {P.}~\bibnamefont
  {Illien}}, \bibinfo {author} {\bibfnamefont {X.}~\bibnamefont {Zhao}},
  \bibinfo {author} {\bibfnamefont {K.~K.}\ \bibnamefont {Dey}}, \bibinfo
  {author} {\bibfnamefont {P.~J.}\ \bibnamefont {Butler}}, \bibinfo {author}
  {\bibfnamefont {A.}~\bibnamefont {Sen}},\ and\ \bibinfo {author}
  {\bibfnamefont {R.}~\bibnamefont {Golestanian}},\ }\bibfield  {title}
  {\enquote {\bibinfo {title} {Exothermicity is not a necessary condition for
  enhanced diffusion of enzymes},}\ }\href
  {https://doi.org/10.1021/acs.nanolett.7b01502} {\bibfield  {journal}
  {\bibinfo  {journal} {Nano Lett.}\ }\textbf {\bibinfo {volume} {17}},\
  \bibinfo {pages} {4415--4420} (\bibinfo {year} {2017})}\BibitemShut {NoStop}%
\bibitem [{\citenamefont {Agudo-Canalejo}, \citenamefont {Illien},\ and\
  \citenamefont {Golestanian}(2018)}]{Agudo-Canalejo:2018a}%
  \BibitemOpen
  \bibfield  {author} {\bibinfo {author} {\bibfnamefont {J.}~\bibnamefont
  {Agudo-Canalejo}}, \bibinfo {author} {\bibfnamefont {P.}~\bibnamefont
  {Illien}},\ and\ \bibinfo {author} {\bibfnamefont {R.}~\bibnamefont
  {Golestanian}},\ }\bibfield  {title} {\enquote {\bibinfo {title} {Phoresis
  and enhanced diffusion compete in enzyme chemotaxis},}\ }\href
  {https://doi.org/10.1021/acs.nanolett.8b00717} {\bibfield  {journal}
  {\bibinfo  {journal} {Nano Lett.}\ }\textbf {\bibinfo {volume} {18}},\
  \bibinfo {pages} {2711--2717} (\bibinfo {year} {2018})}\BibitemShut {NoStop}%
\bibitem [{\citenamefont {Agudo-Canalejo}\ \emph {et~al.}(2018)\citenamefont
  {Agudo-Canalejo}, \citenamefont {Adeleke-Larodo}, \citenamefont {Illien},\
  and\ \citenamefont {Golestanian}}]{Agudo-Canalejo/etal:2018}%
  \BibitemOpen
  \bibfield  {author} {\bibinfo {author} {\bibfnamefont {J.}~\bibnamefont
  {Agudo-Canalejo}}, \bibinfo {author} {\bibfnamefont {T.}~\bibnamefont
  {Adeleke-Larodo}}, \bibinfo {author} {\bibfnamefont {P.}~\bibnamefont
  {Illien}},\ and\ \bibinfo {author} {\bibfnamefont {R.}~\bibnamefont
  {Golestanian}},\ }\bibfield  {title} {\enquote {\bibinfo {title} {Enhanced
  diffusion and chemotaxis at the nanoscale},}\ }\href
  {https://doi.org/10.1021/acs.accounts.8b00280} {\bibfield  {journal}
  {\bibinfo  {journal} {Acc. Chem. Res.}\ }\textbf {\bibinfo {volume} {51}},\
  \bibinfo {pages} {2365} (\bibinfo {year} {2018})}\BibitemShut {NoStop}%
\bibitem [{\citenamefont {Kondrat}\ and\ \citenamefont
  {Popescu}(2019)}]{Kondrat/Popescu:2019}%
  \BibitemOpen
  \bibfield  {author} {\bibinfo {author} {\bibfnamefont {S.}~\bibnamefont
  {Kondrat}}\ and\ \bibinfo {author} {\bibfnamefont {M.~N.}\ \bibnamefont
  {Popescu}},\ }\bibfield  {title} {\enquote {\bibinfo {title} {Brownian
  dynamics assessment of enhanced diffusion exhibited by
  ‘fluctuating-dumbbell enzymes’},}\ }\href
  {https://doi.org/10.1039/C9CP02842K} {\bibfield  {journal} {\bibinfo
  {journal} {Phys. Chem. Chem. Phys.}\ }\textbf {\bibinfo {volume} {21}},\
  \bibinfo {pages} {18811--18815} (\bibinfo {year} {2019})}\BibitemShut
  {NoStop}%
\bibitem [{\citenamefont {Wang}, \citenamefont {Huang},\ and\ \citenamefont
  {Granick}(2021)}]{Huan/etal:JPCLett2021}%
  \BibitemOpen
  \bibfield  {author} {\bibinfo {author} {\bibfnamefont {H.}~\bibnamefont
  {Wang}}, \bibinfo {author} {\bibfnamefont {T.}~\bibnamefont {Huang}},\ and\
  \bibinfo {author} {\bibfnamefont {S.}~\bibnamefont {Granick}},\ }\bibfield
  {title} {\enquote {\bibinfo {title} {Using {NMR} to test molecular mobility
  during a chemical reaction},}\ }\href
  {https://doi.org/10.1021/acs.jpclett.1c00066} {\bibfield  {journal} {\bibinfo
   {journal} {J. Phys. Chem. Lett.}\ }\textbf {\bibinfo {volume} {12}},\
  \bibinfo {pages} {2370--2375} (\bibinfo {year} {2021})}\BibitemShut {NoStop}%
\bibitem [{\citenamefont {G{\" u}nther}, \citenamefont {Majer},\ and\
  \citenamefont {Fischer}(2019)}]{Guenther/etal:JCP2019}%
  \BibitemOpen
  \bibfield  {author} {\bibinfo {author} {\bibfnamefont {J.-P.}\ \bibnamefont
  {G{\" u}nther}}, \bibinfo {author} {\bibfnamefont {G.}~\bibnamefont
  {Majer}},\ and\ \bibinfo {author} {\bibfnamefont {P.}~\bibnamefont
  {Fischer}},\ }\bibfield  {title} {\enquote {\bibinfo {title} {Absolute
  diffusion measurements of active enzyme solutions by {NMR}},}\ }\href
  {https://doi.org/10.1063/1.5086427} {\bibfield  {journal} {\bibinfo
  {journal} {J. Chem. Phys.}\ }\textbf {\bibinfo {volume} {150}},\ \bibinfo
  {pages} {124201} (\bibinfo {year} {2019})}\BibitemShut {NoStop}%
\bibitem [{\citenamefont {Wang}\ \emph {et~al.}(2020)\citenamefont {Wang},
  \citenamefont {Park}, \citenamefont {Dong}, \citenamefont {Kim},
  \citenamefont {Cho}, \citenamefont {Tlusty},\ and\ \citenamefont
  {Granick}}]{Wang/etal:2020}%
  \BibitemOpen
  \bibfield  {author} {\bibinfo {author} {\bibfnamefont {H.}~\bibnamefont
  {Wang}}, \bibinfo {author} {\bibfnamefont {M.}~\bibnamefont {Park}}, \bibinfo
  {author} {\bibfnamefont {R.}~\bibnamefont {Dong}}, \bibinfo {author}
  {\bibfnamefont {J.}~\bibnamefont {Kim}}, \bibinfo {author} {\bibfnamefont
  {Y.-K.}\ \bibnamefont {Cho}}, \bibinfo {author} {\bibfnamefont
  {T.}~\bibnamefont {Tlusty}},\ and\ \bibinfo {author} {\bibfnamefont
  {S.}~\bibnamefont {Granick}},\ }\bibfield  {title} {\enquote {\bibinfo
  {title} {Boosted molecular mobility during common chemical reactions},}\
  }\href {https://doi.org/10.1126/science.aba8425} {\bibfield  {journal}
  {\bibinfo  {journal} {Science}\ }\textbf {\bibinfo {volume} {369}},\ \bibinfo
  {pages} {537--541} (\bibinfo {year} {2020})}\BibitemShut {NoStop}%
\bibitem [{\citenamefont {G{\"u}nther}\ \emph {et~al.}(2021)\citenamefont
  {G{\"u}nther}, \citenamefont {Fillbrook}, \citenamefont {MacDonald},
  \citenamefont {Majer}, \citenamefont {Price}, \citenamefont {Fischer},\ and\
  \citenamefont {Beves}}]{Gunther/etal:2020}%
  \BibitemOpen
  \bibfield  {author} {\bibinfo {author} {\bibfnamefont {J.-P.}\ \bibnamefont
  {G{\"u}nther}}, \bibinfo {author} {\bibfnamefont {L.~L.}\ \bibnamefont
  {Fillbrook}}, \bibinfo {author} {\bibfnamefont {T.~S.~C.}\ \bibnamefont
  {MacDonald}}, \bibinfo {author} {\bibfnamefont {G.}~\bibnamefont {Majer}},
  \bibinfo {author} {\bibfnamefont {W.~S.}\ \bibnamefont {Price}}, \bibinfo
  {author} {\bibfnamefont {P.}~\bibnamefont {Fischer}},\ and\ \bibinfo {author}
  {\bibfnamefont {J.~E.}\ \bibnamefont {Beves}},\ }\bibfield  {title} {\enquote
  {\bibinfo {title} {Comment on {\textquotedblleft}{Boosted} molecular mobility
  during common chemical reactions{\textquotedblright}},}\ }\href
  {https://doi.org/10.1126/science.abe8322} {\bibfield  {journal} {\bibinfo
  {journal} {Science}\ }\textbf {\bibinfo {volume} {371}},\ \bibinfo {pages}
  {eabe8322} (\bibinfo {year} {2021})}\BibitemShut {NoStop}%
\bibitem [{\citenamefont {Wang}\ \emph {et~al.}(2021)\citenamefont {Wang},
  \citenamefont {Park}, \citenamefont {Dong}, \citenamefont {Kim},
  \citenamefont {Cho}, \citenamefont {Tlusty},\ and\ \citenamefont
  {Granick}}]{Wang/etal:2020:2}%
  \BibitemOpen
  \bibfield  {author} {\bibinfo {author} {\bibfnamefont {H.}~\bibnamefont
  {Wang}}, \bibinfo {author} {\bibfnamefont {M.}~\bibnamefont {Park}}, \bibinfo
  {author} {\bibfnamefont {R.}~\bibnamefont {Dong}}, \bibinfo {author}
  {\bibfnamefont {J.}~\bibnamefont {Kim}}, \bibinfo {author} {\bibfnamefont
  {Y.-K.}\ \bibnamefont {Cho}}, \bibinfo {author} {\bibfnamefont
  {T.}~\bibnamefont {Tlusty}},\ and\ \bibinfo {author} {\bibfnamefont
  {S.}~\bibnamefont {Granick}},\ }\bibfield  {title} {\enquote {\bibinfo
  {title} {Response to comment on {\textquotedblleft}{Boosted} molecular
  mobility during common chemical reactions{\textquotedblright}},}\ }\href
  {https://doi.org/10.1126/science.abe8678} {\bibfield  {journal} {\bibinfo
  {journal} {Science}\ }\textbf {\bibinfo {volume} {371}},\ \bibinfo {pages}
  {eabe8678} (\bibinfo {year} {2021})}\BibitemShut {NoStop}%
\bibitem [{\citenamefont {Feng}\ and\ \citenamefont
  {Gilson}(2020)}]{Feng:2020}%
  \BibitemOpen
  \bibfield  {author} {\bibinfo {author} {\bibfnamefont {M.}~\bibnamefont
  {Feng}}\ and\ \bibinfo {author} {\bibfnamefont {M.~K.}\ \bibnamefont
  {Gilson}},\ }\bibfield  {title} {\enquote {\bibinfo {title} {Enhanced
  diffusion and chemotaxis of enzymes},}\ }\href
  {https://doi.org/10.1146/annurev-biophys-121219-081535} {\bibfield  {journal}
  {\bibinfo  {journal} {Annu. Rev. Biophys}\ }\textbf {\bibinfo {volume}
  {49}},\ \bibinfo {pages} {87--105} (\bibinfo {year} {2020})}\BibitemShut
  {NoStop}%
\bibitem [{\citenamefont {Tolman}(1925)}]{Tolman:PNAS1925}%
  \BibitemOpen
  \bibfield  {author} {\bibinfo {author} {\bibfnamefont {R.~C.}\ \bibnamefont
  {Tolman}},\ }\bibfield  {title} {\enquote {\bibinfo {title} {The principle of
  microscopic reversibility},}\ }\href {https://doi.org/10.1073/pnas.11.7.436}
  {\bibfield  {journal} {\bibinfo  {journal} {Proc. Natl. Acad. Sci. U.S.A.}\
  }\textbf {\bibinfo {volume} {11}},\ \bibinfo {pages} {436--439} (\bibinfo
  {year} {1925})}\BibitemShut {NoStop}%
\bibitem [{\citenamefont {Ryabov}\ and\ \citenamefont
  {Tasinkevych}(2022)}]{Ryabov/Tasinkevych:SoftMatt2022}%
  \BibitemOpen
  \bibfield  {author} {\bibinfo {author} {\bibfnamefont {A.}~\bibnamefont
  {Ryabov}}\ and\ \bibinfo {author} {\bibfnamefont {M.}~\bibnamefont
  {Tasinkevych}},\ }\bibfield  {title} {\enquote {\bibinfo {title} {Enhanced
  diffusivity in microscopically reversible active matter},}\ }\href
  {https://doi.org/10.1039/D2SM00054G} {\bibfield  {journal} {\bibinfo
  {journal} {Soft Matter}\ }\textbf {\bibinfo {volume} {18}},\ \bibinfo {pages}
  {3234--3240} (\bibinfo {year} {2022})}\BibitemShut {NoStop}%
\bibitem [{\citenamefont {Han}\ \emph {et~al.}(2006)\citenamefont {Han},
  \citenamefont {Alsayed}, \citenamefont {Nobili}, \citenamefont {Zhang},
  \citenamefont {Lubensky},\ and\ \citenamefont {Yodh}}]{Han/etal:SCIENCE2006}%
  \BibitemOpen
  \bibfield  {author} {\bibinfo {author} {\bibfnamefont {Y.}~\bibnamefont
  {Han}}, \bibinfo {author} {\bibfnamefont {A.~M.}\ \bibnamefont {Alsayed}},
  \bibinfo {author} {\bibfnamefont {M.}~\bibnamefont {Nobili}}, \bibinfo
  {author} {\bibfnamefont {J.}~\bibnamefont {Zhang}}, \bibinfo {author}
  {\bibfnamefont {T.~C.}\ \bibnamefont {Lubensky}},\ and\ \bibinfo {author}
  {\bibfnamefont {A.~G.}\ \bibnamefont {Yodh}},\ }\bibfield  {title} {\enquote
  {\bibinfo {title} {Brownian motion of an ellipsoid},}\ }\href
  {https://doi.org/10.1126/science.1130146} {\bibfield  {journal} {\bibinfo
  {journal} {Science}\ }\textbf {\bibinfo {volume} {314}},\ \bibinfo {pages}
  {626--630} (\bibinfo {year} {2006})}\BibitemShut {NoStop}%
\bibitem [{\citenamefont {Onsager}(1931)}]{Onsager:1931a}%
  \BibitemOpen
  \bibfield  {author} {\bibinfo {author} {\bibfnamefont {L.}~\bibnamefont
  {Onsager}},\ }\bibfield  {title} {\enquote {\bibinfo {title} {Reciprocal
  relations in irreversible processes. {I.}}}\ }\href
  {https://doi.org/10.1103/PhysRev.37.405} {\bibfield  {journal} {\bibinfo
  {journal} {Phys. Rev.}\ }\textbf {\bibinfo {volume} {37}},\ \bibinfo {pages}
  {405--426} (\bibinfo {year} {1931})}\BibitemShut {NoStop}%
\bibitem [{\citenamefont {Onsager}\ and\ \citenamefont
  {Machlup}(1953)}]{Onsager/Machlup:1953}%
  \BibitemOpen
  \bibfield  {author} {\bibinfo {author} {\bibfnamefont {L.}~\bibnamefont
  {Onsager}}\ and\ \bibinfo {author} {\bibfnamefont {S.}~\bibnamefont
  {Machlup}},\ }\bibfield  {title} {\enquote {\bibinfo {title} {Fluctuations
  and irreversible processes},}\ }\href
  {https://doi.org/10.1103/PhysRev.91.1505} {\bibfield  {journal} {\bibinfo
  {journal} {Phys. Rev.}\ }\textbf {\bibinfo {volume} {91}},\ \bibinfo {pages}
  {1505} (\bibinfo {year} {1953})}\BibitemShut {NoStop}%
\bibitem [{\citenamefont {Astumian}(2016)}]{Astumian:2016}%
  \BibitemOpen
  \bibfield  {author} {\bibinfo {author} {\bibfnamefont {R.~D.}\ \bibnamefont
  {Astumian}},\ }\bibfield  {title} {\enquote {\bibinfo {title} {Optical vs.
  chemical driving for molecular machines},}\ }\href
  {https://doi.org/10.1039/C6FD00140H} {\bibfield  {journal} {\bibinfo
  {journal} {Faraday Discuss.}\ }\textbf {\bibinfo {volume} {195}},\ \bibinfo
  {pages} {583--597} (\bibinfo {year} {2016})}\BibitemShut {NoStop}%
\bibitem [{\citenamefont {Maes}(2021)}]{Maes:SciPost2021}%
  \BibitemOpen
  \bibfield  {author} {\bibinfo {author} {\bibfnamefont {C.}~\bibnamefont
  {Maes}},\ }\bibfield  {title} {\enquote {\bibinfo {title} {{Local detailed
  balance}},}\ }\href {https://doi.org/10.21468/SciPostPhysLectNotes.32}
  {\bibfield  {journal} {\bibinfo  {journal} {SciPost Phys. Lect. Notes}\ ,\
  \bibinfo {pages} {32}} (\bibinfo {year} {2021})}\BibitemShut {NoStop}%
\bibitem [{\citenamefont {Zwanzig}(2001)}]{Zwanzig:book2001}%
  \BibitemOpen
  \bibfield  {author} {\bibinfo {author} {\bibfnamefont {R.}~\bibnamefont
  {Zwanzig}},\ }\href@noop {} {\emph {\bibinfo {title} {Nonequilibrium
  statistical mechanics}}}\ (\bibinfo  {publisher} {Oxford University Press,
  Oxford},\ \bibinfo {year} {2001})\BibitemShut {NoStop}%
\bibitem [{\citenamefont {Metzler}\ \emph {et~al.}(2014)\citenamefont
  {Metzler}, \citenamefont {Jeon}, \citenamefont {Cherstvy},\ and\
  \citenamefont {Barkai}}]{Metzler/etal:PCCP2014}%
  \BibitemOpen
  \bibfield  {author} {\bibinfo {author} {\bibfnamefont {R.}~\bibnamefont
  {Metzler}}, \bibinfo {author} {\bibfnamefont {J.-H.}\ \bibnamefont {Jeon}},
  \bibinfo {author} {\bibfnamefont {A.~G.}\ \bibnamefont {Cherstvy}},\ and\
  \bibinfo {author} {\bibfnamefont {E.}~\bibnamefont {Barkai}},\ }\bibfield
  {title} {\enquote {\bibinfo {title} {Anomalous diffusion models and their
  properties: non-stationarity{,} non-ergodicity{,} and ageing at the centenary
  of single particle tracking},}\ }\href {https://doi.org/10.1039/C4CP03465A}
  {\bibfield  {journal} {\bibinfo  {journal} {Phys. Chem. Chem. Phys.}\
  }\textbf {\bibinfo {volume} {16}},\ \bibinfo {pages} {24128--24164} (\bibinfo
  {year} {2014})}\BibitemShut {NoStop}%
\bibitem [{\citenamefont {Chen}\ \emph {et~al.}(2020)\citenamefont {Chen},
  \citenamefont {Shaw}, \citenamefont {Wilson}, \citenamefont {Woringer},
  \citenamefont {Darzacq}, \citenamefont {Marqusee}, \citenamefont {Wang},\
  and\ \citenamefont {Bustamante}}]{Chen/etal:PNAS2020}%
  \BibitemOpen
  \bibfield  {author} {\bibinfo {author} {\bibfnamefont {Z.}~\bibnamefont
  {Chen}}, \bibinfo {author} {\bibfnamefont {A.}~\bibnamefont {Shaw}}, \bibinfo
  {author} {\bibfnamefont {H.}~\bibnamefont {Wilson}}, \bibinfo {author}
  {\bibfnamefont {M.}~\bibnamefont {Woringer}}, \bibinfo {author}
  {\bibfnamefont {X.}~\bibnamefont {Darzacq}}, \bibinfo {author} {\bibfnamefont
  {S.}~\bibnamefont {Marqusee}}, \bibinfo {author} {\bibfnamefont
  {Q.}~\bibnamefont {Wang}},\ and\ \bibinfo {author} {\bibfnamefont
  {C.}~\bibnamefont {Bustamante}},\ }\bibfield  {title} {\enquote {\bibinfo
  {title} {Single-molecule diffusometry reveals no catalysis-induced diffusion
  enhancement of alkaline phosphatase as proposed by {FCS} experiments},}\
  }\href {https://doi.org/10.1073/pnas.2006900117} {\bibfield  {journal}
  {\bibinfo  {journal} {Proc. Natl. Acad. Sci. U.S.A.}\ }\textbf {\bibinfo
  {volume} {117}},\ \bibinfo {pages} {21328--21335} (\bibinfo {year}
  {2020})}\BibitemShut {NoStop}%
\bibitem [{\citenamefont {Evans}(2020)}]{Evans:2020}%
  \BibitemOpen
  \bibfield  {author} {\bibinfo {author} {\bibfnamefont {R.}~\bibnamefont
  {Evans}},\ }\bibfield  {title} {\enquote {\bibinfo {title} {The
  interpretation of small molecule diffusion coefficients: Quantitative use of
  diffusion-ordered {NMR} spectroscopy},}\ }\href
  {https://doi.org/https://doi.org/10.1016/j.pnmrs.2019.11.002} {\bibfield
  {journal} {\bibinfo  {journal} {Prog. Nucl. Magn. Reson. Spectrosc.}\
  }\textbf {\bibinfo {volume} {117}},\ \bibinfo {pages} {33--69} (\bibinfo
  {year} {2020})}\BibitemShut {NoStop}%
\bibitem [{\citenamefont {K{\"a}rger}\ \emph {et~al.}(2021)\citenamefont
  {K{\"a}rger}, \citenamefont {Avramovska}, \citenamefont {Freude},
  \citenamefont {Haase}, \citenamefont {Hwang},\ and\ \citenamefont
  {Valiullin}}]{Kaerger/etal:2021}%
  \BibitemOpen
  \bibfield  {author} {\bibinfo {author} {\bibfnamefont {J.}~\bibnamefont
  {K{\"a}rger}}, \bibinfo {author} {\bibfnamefont {M.}~\bibnamefont
  {Avramovska}}, \bibinfo {author} {\bibfnamefont {D.}~\bibnamefont {Freude}},
  \bibinfo {author} {\bibfnamefont {J.}~\bibnamefont {Haase}}, \bibinfo
  {author} {\bibfnamefont {S.}~\bibnamefont {Hwang}},\ and\ \bibinfo {author}
  {\bibfnamefont {R.}~\bibnamefont {Valiullin}},\ }\bibfield  {title} {\enquote
  {\bibinfo {title} {Pulsed field gradient {NMR} diffusion measurement in
  nanoporous materials},}\ }\href {https://doi.org/10.1007/s10450-020-00290-9}
  {\bibfield  {journal} {\bibinfo  {journal} {Adsorption}\ }\textbf {\bibinfo
  {volume} {27}},\ \bibinfo {pages} {453--484} (\bibinfo {year}
  {2021})}\BibitemShut {NoStop}%
\bibitem [{\citenamefont {Jobic}\ and\ \citenamefont
  {Theodorou}(2007)}]{Jobic/Theodorou:2007}%
  \BibitemOpen
  \bibfield  {author} {\bibinfo {author} {\bibfnamefont {H.}~\bibnamefont
  {Jobic}}\ and\ \bibinfo {author} {\bibfnamefont {D.~N.}\ \bibnamefont
  {Theodorou}},\ }\bibfield  {title} {\enquote {\bibinfo {title} {Quasi-elastic
  neutron scattering and molecular dynamics simulation as complementary
  techniques for studying diffusion in zeolites},}\ }\href
  {https://doi.org/https://doi.org/10.1016/j.micromeso.2006.12.034} {\bibfield
  {journal} {\bibinfo  {journal} {Micropor. Mesopor. Mat.}\ }\textbf {\bibinfo
  {volume} {102}},\ \bibinfo {pages} {21--50} (\bibinfo {year}
  {2007})}\BibitemShut {NoStop}%
\bibitem [{\citenamefont {Ryabov}, \citenamefont {{\v Z}onda},\ and\
  \citenamefont {Novotn{\' y}}(2022)}]{Ryabov/etal:CNSNS2022}%
  \BibitemOpen
  \bibfield  {author} {\bibinfo {author} {\bibfnamefont {A.}~\bibnamefont
  {Ryabov}}, \bibinfo {author} {\bibfnamefont {M.}~\bibnamefont {{\v Z}onda}},\
  and\ \bibinfo {author} {\bibfnamefont {T.}~\bibnamefont {Novotn{\' y}}},\
  }\bibfield  {title} {\enquote {\bibinfo {title} {Phase diffusion and noise
  temperature of a microwave amplifier based on single unshunted {Josephson}
  junction},}\ }\href
  {https://doi.org/https://doi.org/10.1016/j.cnsns.2022.106523} {\bibfield
  {journal} {\bibinfo  {journal} {Comm. Nonlinear Sci. Numer. Simulat.}\
  }\textbf {\bibinfo {volume} {112}},\ \bibinfo {pages} {106523} (\bibinfo
  {year} {2022})}\BibitemShut {NoStop}%
\bibitem [{\citenamefont {Erdmann}\ \emph {et~al.}(2000)\citenamefont
  {Erdmann}, \citenamefont {Ebeling}, \citenamefont {Schimansky-Geier},\ and\
  \citenamefont {Schweitzer}}]{Erdmann/etal:EPJB2000}%
  \BibitemOpen
  \bibfield  {author} {\bibinfo {author} {\bibfnamefont {U.}~\bibnamefont
  {Erdmann}}, \bibinfo {author} {\bibfnamefont {W.}~\bibnamefont {Ebeling}},
  \bibinfo {author} {\bibfnamefont {L.}~\bibnamefont {Schimansky-Geier}},\ and\
  \bibinfo {author} {\bibfnamefont {F.}~\bibnamefont {Schweitzer}},\ }\bibfield
   {title} {\enquote {\bibinfo {title} {Brownian particles far from
  equilibrium},}\ }\href {https://doi.org/10.1007/s100510051104} {\bibfield
  {journal} {\bibinfo  {journal} {Eur. Phys. J. B}\ }\textbf {\bibinfo {volume}
  {15}},\ \bibinfo {pages} {105--113} (\bibinfo {year} {2000})}\BibitemShut
  {NoStop}%
\bibitem [{\citenamefont {Szab\'o}\ \emph {et~al.}(2006)\citenamefont
  {Szab\'o}, \citenamefont {Sz\"oll\"osi}, \citenamefont {G\"onci},
  \citenamefont {Jur\'anyi}, \citenamefont {Selmeczi},\ and\ \citenamefont
  {Vicsek}}]{Szabo/etal:PRE2006}%
  \BibitemOpen
  \bibfield  {author} {\bibinfo {author} {\bibfnamefont {B.}~\bibnamefont
  {Szab\'o}}, \bibinfo {author} {\bibfnamefont {G.~J.}\ \bibnamefont
  {Sz\"oll\"osi}}, \bibinfo {author} {\bibfnamefont {B.}~\bibnamefont
  {G\"onci}}, \bibinfo {author} {\bibfnamefont {Z.}~\bibnamefont {Jur\'anyi}},
  \bibinfo {author} {\bibfnamefont {D.}~\bibnamefont {Selmeczi}},\ and\
  \bibinfo {author} {\bibfnamefont {T.}~\bibnamefont {Vicsek}},\ }\bibfield
  {title} {\enquote {\bibinfo {title} {Phase transition in the collective
  migration of tissue cells: Experiment and model},}\ }\href
  {https://doi.org/10.1103/PhysRevE.74.061908} {\bibfield  {journal} {\bibinfo
  {journal} {Phys. Rev. E}\ }\textbf {\bibinfo {volume} {74}},\ \bibinfo
  {pages} {061908} (\bibinfo {year} {2006})}\BibitemShut {NoStop}%
\bibitem [{\citenamefont {Peruani}\ and\ \citenamefont
  {Morelli}(2007)}]{Peruani/Morelli:PRL2007}%
  \BibitemOpen
  \bibfield  {author} {\bibinfo {author} {\bibfnamefont {F.}~\bibnamefont
  {Peruani}}\ and\ \bibinfo {author} {\bibfnamefont {L.~G.}\ \bibnamefont
  {Morelli}},\ }\bibfield  {title} {\enquote {\bibinfo {title} {Self-propelled
  particles with fluctuating speed and direction of motion in two
  dimensions},}\ }\href {https://doi.org/10.1103/PhysRevLett.99.010602}
  {\bibfield  {journal} {\bibinfo  {journal} {Phys. Rev. Lett.}\ }\textbf
  {\bibinfo {volume} {99}},\ \bibinfo {pages} {010602} (\bibinfo {year}
  {2007})}\BibitemShut {NoStop}%
\bibitem [{\citenamefont {van Teeffelen}\ and\ \citenamefont
  {L\"owen}(2008)}]{Teeffelen/Loewen:PRE2008}%
  \BibitemOpen
  \bibfield  {author} {\bibinfo {author} {\bibfnamefont {S.}~\bibnamefont {van
  Teeffelen}}\ and\ \bibinfo {author} {\bibfnamefont {H.}~\bibnamefont
  {L\"owen}},\ }\bibfield  {title} {\enquote {\bibinfo {title} {Dynamics of a
  {Brownian} circle swimmer},}\ }\href
  {https://doi.org/10.1103/PhysRevE.78.020101} {\bibfield  {journal} {\bibinfo
  {journal} {Phys. Rev. E}\ }\textbf {\bibinfo {volume} {78}},\ \bibinfo
  {pages} {020101} (\bibinfo {year} {2008})}\BibitemShut {NoStop}%
\bibitem [{\citenamefont {{ten Hagen}}, \citenamefont {{van Teeffelen}},\ and\
  \citenamefont {L\"owen}(2011)}]{tenHagen/etal:2011}%
  \BibitemOpen
  \bibfield  {author} {\bibinfo {author} {\bibfnamefont {B.}~\bibnamefont {{ten
  Hagen}}}, \bibinfo {author} {\bibfnamefont {S.}~\bibnamefont {{van
  Teeffelen}}},\ and\ \bibinfo {author} {\bibfnamefont {H.}~\bibnamefont
  {L\"owen}},\ }\bibfield  {title} {\enquote {\bibinfo {title} {Brownian motion
  of a self-propelled particle},}\ }\href
  {https://doi.org/10.1088/0953-8984/23/19/194119} {\bibfield  {journal}
  {\bibinfo  {journal} {J. Phys.: Condens. Matt.}\ }\textbf {\bibinfo {volume}
  {23}},\ \bibinfo {pages} {194119} (\bibinfo {year} {2011})}\BibitemShut
  {NoStop}%
\bibitem [{\citenamefont {Henkes}, \citenamefont {Fily},\ and\ \citenamefont
  {Marchetti}(2011)}]{Henkes/etal:2011}%
  \BibitemOpen
  \bibfield  {author} {\bibinfo {author} {\bibfnamefont {S.}~\bibnamefont
  {Henkes}}, \bibinfo {author} {\bibfnamefont {Y.}~\bibnamefont {Fily}},\ and\
  \bibinfo {author} {\bibfnamefont {M.~C.}\ \bibnamefont {Marchetti}},\
  }\bibfield  {title} {\enquote {\bibinfo {title} {Active jamming:
  Self-propelled soft particles at high density},}\ }\href
  {https://doi.org/10.1103/PhysRevE.84.040301} {\bibfield  {journal} {\bibinfo
  {journal} {Phys. Rev. E}\ }\textbf {\bibinfo {volume} {84}},\ \bibinfo
  {pages} {040301} (\bibinfo {year} {2011})}\BibitemShut {NoStop}%
\bibitem [{\citenamefont {Bialk\'e}, \citenamefont {Speck},\ and\ \citenamefont
  {L\"owen}(2012)}]{Bialke/etal:PRL2012}%
  \BibitemOpen
  \bibfield  {author} {\bibinfo {author} {\bibfnamefont {J.}~\bibnamefont
  {Bialk\'e}}, \bibinfo {author} {\bibfnamefont {T.}~\bibnamefont {Speck}},\
  and\ \bibinfo {author} {\bibfnamefont {H.}~\bibnamefont {L\"owen}},\
  }\bibfield  {title} {\enquote {\bibinfo {title} {Crystallization in a dense
  suspension of self-propelled particles},}\ }\href
  {https://doi.org/10.1103/PhysRevLett.108.168301} {\bibfield  {journal}
  {\bibinfo  {journal} {Phys. Rev. Lett.}\ }\textbf {\bibinfo {volume} {108}},\
  \bibinfo {pages} {168301} (\bibinfo {year} {2012})}\BibitemShut {NoStop}%
\bibitem [{\citenamefont {Romanczuk}\ \emph {et~al.}(2012)\citenamefont
  {Romanczuk}, \citenamefont {B{\"a}r}, \citenamefont {Ebeling}, \citenamefont
  {Lindner},\ and\ \citenamefont {Schimansky-Geier}}]{Romanczuk2012}%
  \BibitemOpen
  \bibfield  {author} {\bibinfo {author} {\bibfnamefont {P.}~\bibnamefont
  {Romanczuk}}, \bibinfo {author} {\bibfnamefont {M.}~\bibnamefont {B{\"a}r}},
  \bibinfo {author} {\bibfnamefont {W.}~\bibnamefont {Ebeling}}, \bibinfo
  {author} {\bibfnamefont {B.}~\bibnamefont {Lindner}},\ and\ \bibinfo {author}
  {\bibfnamefont {L.}~\bibnamefont {Schimansky-Geier}},\ }\bibfield  {title}
  {\enquote {\bibinfo {title} {Active {Brownian} particles},}\ }\href
  {https://doi.org/10.1140/epjst/e2012-01529-y} {\bibfield  {journal} {\bibinfo
   {journal} {Eur. Phys. J. Spec. Top.}\ }\textbf {\bibinfo {volume} {202}},\
  \bibinfo {pages} {1--162} (\bibinfo {year} {2012})}\BibitemShut {NoStop}%
\bibitem [{\citenamefont {Pototsky}\ and\ \citenamefont
  {Stark}(2012)}]{Pototsky/Stark:EPL2012}%
  \BibitemOpen
  \bibfield  {author} {\bibinfo {author} {\bibfnamefont {A.}~\bibnamefont
  {Pototsky}}\ and\ \bibinfo {author} {\bibfnamefont {H.}~\bibnamefont
  {Stark}},\ }\bibfield  {title} {\enquote {\bibinfo {title} {Active {Brownian}
  particles in two-dimensional traps},}\ }\href
  {https://doi.org/10.1209/0295-5075/98/50004} {\bibfield  {journal} {\bibinfo
  {journal} {{EPL} (Europhysics Letters)}\ }\textbf {\bibinfo {volume} {98}},\
  \bibinfo {pages} {50004} (\bibinfo {year} {2012})}\BibitemShut {NoStop}%
\bibitem [{\citenamefont {Buttinoni}\ \emph {et~al.}(2013)\citenamefont
  {Buttinoni}, \citenamefont {Bialk\'e}, \citenamefont {K\"ummel},
  \citenamefont {L\"owen}, \citenamefont {Bechinger},\ and\ \citenamefont
  {Speck}}]{Buttinoni/etal:2013}%
  \BibitemOpen
  \bibfield  {author} {\bibinfo {author} {\bibfnamefont {I.}~\bibnamefont
  {Buttinoni}}, \bibinfo {author} {\bibfnamefont {J.}~\bibnamefont {Bialk\'e}},
  \bibinfo {author} {\bibfnamefont {F.}~\bibnamefont {K\"ummel}}, \bibinfo
  {author} {\bibfnamefont {H.}~\bibnamefont {L\"owen}}, \bibinfo {author}
  {\bibfnamefont {C.}~\bibnamefont {Bechinger}},\ and\ \bibinfo {author}
  {\bibfnamefont {T.}~\bibnamefont {Speck}},\ }\bibfield  {title} {\enquote
  {\bibinfo {title} {Dynamical clustering and phase separation in suspensions
  of self-propelled colloidal particles},}\ }\href
  {https://doi.org/10.1103/PhysRevLett.110.238301} {\bibfield  {journal}
  {\bibinfo  {journal} {Phys. Rev. Lett.}\ }\textbf {\bibinfo {volume} {110}},\
  \bibinfo {pages} {238301} (\bibinfo {year} {2013})}\BibitemShut {NoStop}%
\bibitem [{\citenamefont {Yang}, \citenamefont {Manning},\ and\ \citenamefont
  {Marchetti}(2014)}]{Yang/etal:SOFTMATTER2014}%
  \BibitemOpen
  \bibfield  {author} {\bibinfo {author} {\bibfnamefont {X.}~\bibnamefont
  {Yang}}, \bibinfo {author} {\bibfnamefont {M.~L.}\ \bibnamefont {Manning}},\
  and\ \bibinfo {author} {\bibfnamefont {M.~C.}\ \bibnamefont {Marchetti}},\
  }\bibfield  {title} {\enquote {\bibinfo {title} {Aggregation and segregation
  of confined active particles},}\ }\href {https://doi.org/10.1039/C4SM00927D}
  {\bibfield  {journal} {\bibinfo  {journal} {Soft Matter}\ }\textbf {\bibinfo
  {volume} {10}},\ \bibinfo {pages} {6477--6484} (\bibinfo {year}
  {2014})}\BibitemShut {NoStop}%
\bibitem [{\citenamefont {Stenhammar}\ \emph {et~al.}(2014)\citenamefont
  {Stenhammar}, \citenamefont {Marenduzzo}, \citenamefont {Allen},\ and\
  \citenamefont {Cates}}]{Stenhammar/etal:SOFTMATTER2014}%
  \BibitemOpen
  \bibfield  {author} {\bibinfo {author} {\bibfnamefont {J.}~\bibnamefont
  {Stenhammar}}, \bibinfo {author} {\bibfnamefont {D.}~\bibnamefont
  {Marenduzzo}}, \bibinfo {author} {\bibfnamefont {R.~J.}\ \bibnamefont
  {Allen}},\ and\ \bibinfo {author} {\bibfnamefont {M.~E.}\ \bibnamefont
  {Cates}},\ }\bibfield  {title} {\enquote {\bibinfo {title} {Phase behaviour
  of active brownian particles: the role of dimensionality},}\ }\href
  {https://doi.org/10.1039/C3SM52813H} {\bibfield  {journal} {\bibinfo
  {journal} {Soft Matter}\ }\textbf {\bibinfo {volume} {10}},\ \bibinfo {pages}
  {1489--1499} (\bibinfo {year} {2014})}\BibitemShut {NoStop}%
\bibitem [{\citenamefont {Z\"ottl}\ and\ \citenamefont
  {Stark}(2016)}]{Zottl/Stark:JPCM2016}%
  \BibitemOpen
  \bibfield  {author} {\bibinfo {author} {\bibfnamefont {A.}~\bibnamefont
  {Z\"ottl}}\ and\ \bibinfo {author} {\bibfnamefont {H.}~\bibnamefont
  {Stark}},\ }\bibfield  {title} {\enquote {\bibinfo {title} {Emergent behavior
  in active colloids},}\ }\href
  {https://doi.org/10.1088/0953-8984/28/25/253001} {\bibfield  {journal}
  {\bibinfo  {journal} {J. Phys. Condens. Matt.}\ }\textbf {\bibinfo {volume}
  {28}},\ \bibinfo {pages} {253001} (\bibinfo {year} {2016})}\BibitemShut
  {NoStop}%
\bibitem [{\citenamefont {Das}, \citenamefont {Gompper},\ and\ \citenamefont
  {Winkler}(2018)}]{Das/etal:NJP2018}%
  \BibitemOpen
  \bibfield  {author} {\bibinfo {author} {\bibfnamefont {S.}~\bibnamefont
  {Das}}, \bibinfo {author} {\bibfnamefont {G.}~\bibnamefont {Gompper}},\ and\
  \bibinfo {author} {\bibfnamefont {R.~G.}\ \bibnamefont {Winkler}},\
  }\bibfield  {title} {\enquote {\bibinfo {title} {Confined active {Brownian}
  particles: theoretical description of propulsion-induced accumulation},}\
  }\href {https://doi.org/10.1088/1367-2630/aa9d4b} {\bibfield  {journal}
  {\bibinfo  {journal} {New J. Phys.}\ }\textbf {\bibinfo {volume} {20}},\
  \bibinfo {pages} {015001} (\bibinfo {year} {2018})}\BibitemShut {NoStop}%
\bibitem [{\citenamefont {Malakar}\ \emph {et~al.}(2020)\citenamefont
  {Malakar}, \citenamefont {Das}, \citenamefont {Kundu}, \citenamefont
  {Kumar},\ and\ \citenamefont {Dhar}}]{Malakar/etal:2020}%
  \BibitemOpen
  \bibfield  {author} {\bibinfo {author} {\bibfnamefont {K.}~\bibnamefont
  {Malakar}}, \bibinfo {author} {\bibfnamefont {A.}~\bibnamefont {Das}},
  \bibinfo {author} {\bibfnamefont {A.}~\bibnamefont {Kundu}}, \bibinfo
  {author} {\bibfnamefont {K.~V.}\ \bibnamefont {Kumar}},\ and\ \bibinfo
  {author} {\bibfnamefont {A.}~\bibnamefont {Dhar}},\ }\bibfield  {title}
  {\enquote {\bibinfo {title} {Steady state of an active {Brownian} particle in
  a two-dimensional harmonic trap},}\ }\href
  {https://doi.org/10.1103/PhysRevE.101.022610} {\bibfield  {journal} {\bibinfo
   {journal} {Phys. Rev. E}\ }\textbf {\bibinfo {volume} {101}},\ \bibinfo
  {pages} {022610} (\bibinfo {year} {2020})}\BibitemShut {NoStop}%
\bibitem [{\citenamefont {Chaudhuri}\ and\ \citenamefont
  {Dhar}(2021)}]{Chaudhuri/Dhar:JSTAT2021}%
  \BibitemOpen
  \bibfield  {author} {\bibinfo {author} {\bibfnamefont {D.}~\bibnamefont
  {Chaudhuri}}\ and\ \bibinfo {author} {\bibfnamefont {A.}~\bibnamefont
  {Dhar}},\ }\bibfield  {title} {\enquote {\bibinfo {title} {Active {Brownian}
  particle in harmonic trap: exact computation of moments, and re-entrant
  transition},}\ }\href {https://doi.org/10.1088/1742-5468/abd031} {\bibfield
  {journal} {\bibinfo  {journal} {J. Stat. Mech.}\ }\textbf {\bibinfo {volume}
  {2021}},\ \bibinfo {pages} {013207} (\bibinfo {year} {2021})}\BibitemShut
  {NoStop}%
\bibitem [{\citenamefont {Pietzonka}\ and\ \citenamefont
  {Seifert}(2018)}]{Pietzonka/Seifert:2018}%
  \BibitemOpen
  \bibfield  {author} {\bibinfo {author} {\bibfnamefont {P.}~\bibnamefont
  {Pietzonka}}\ and\ \bibinfo {author} {\bibfnamefont {U.}~\bibnamefont
  {Seifert}},\ }\bibfield  {title} {\enquote {\bibinfo {title} {Entropy
  production of active particles and for particles in active baths},}\ }\href
  {https://doi.org/10.1088/1751-8121/aa91b9} {\bibfield  {journal} {\bibinfo
  {journal} {J. Phys. A}\ }\textbf {\bibinfo {volume} {51}},\ \bibinfo {pages}
  {01LT01} (\bibinfo {year} {2018})}\BibitemShut {NoStop}%
\bibitem [{\citenamefont {Speck}(2018)}]{Speck:2018}%
  \BibitemOpen
  \bibfield  {author} {\bibinfo {author} {\bibfnamefont {T.}~\bibnamefont
  {Speck}},\ }\bibfield  {title} {\enquote {\bibinfo {title} {Active {Brownian}
  particles driven by constant affinity},}\ }\href
  {https://doi.org/10.1209/0295-5075/123/20007} {\bibfield  {journal} {\bibinfo
   {journal} {EPL}\ }\textbf {\bibinfo {volume} {123}},\ \bibinfo {pages}
  {20007} (\bibinfo {year} {2018})}\BibitemShut {NoStop}%
\bibitem [{\citenamefont {Seifert}(2011)}]{Seifert:2011}%
  \BibitemOpen
  \bibfield  {author} {\bibinfo {author} {\bibfnamefont {U.}~\bibnamefont
  {Seifert}},\ }\bibfield  {title} {\enquote {\bibinfo {title} {Stochastic
  thermodynamics of single enzymes and molecular motors},}\ }\href
  {https://doi.org/10.1140/epje/i2011-11026-7} {\bibfield  {journal} {\bibinfo
  {journal} {Eur. Phys. J. E}\ }\textbf {\bibinfo {volume} {34}},\ \bibinfo
  {pages} {26} (\bibinfo {year} {2011})}\BibitemShut {NoStop}%
\bibitem [{\citenamefont {Speck}(2021)}]{Speck:2021}%
  \BibitemOpen
  \bibfield  {author} {\bibinfo {author} {\bibfnamefont {T.}~\bibnamefont
  {Speck}},\ }\bibfield  {title} {\enquote {\bibinfo {title} {Modeling of
  biomolecular machines in non-equilibrium steady states},}\ }\href
  {https://doi.org/10.1063/5.0070922} {\bibfield  {journal} {\bibinfo
  {journal} {J. Chem. Phys.}\ }\textbf {\bibinfo {volume} {155}},\ \bibinfo
  {pages} {230901} (\bibinfo {year} {2021})}\BibitemShut {NoStop}%
\bibitem [{\citenamefont {Speck}(2019)}]{Speck:2019}%
  \BibitemOpen
  \bibfield  {author} {\bibinfo {author} {\bibfnamefont {T.}~\bibnamefont
  {Speck}},\ }\bibfield  {title} {\enquote {\bibinfo {title} {Thermodynamic
  approach to the self-diffusiophoresis of colloidal {Janus} particles},}\
  }\href {https://doi.org/10.1103/PhysRevE.99.060602} {\bibfield  {journal}
  {\bibinfo  {journal} {Phys. Rev. E}\ }\textbf {\bibinfo {volume} {99}},\
  \bibinfo {pages} {060602} (\bibinfo {year} {2019})}\BibitemShut {NoStop}%
\bibitem [{\citenamefont {Fischer}, \citenamefont {Chatterjee},\ and\
  \citenamefont {Speck}(2019)}]{Fisher/etal:2019}%
  \BibitemOpen
  \bibfield  {author} {\bibinfo {author} {\bibfnamefont {A.}~\bibnamefont
  {Fischer}}, \bibinfo {author} {\bibfnamefont {A.}~\bibnamefont
  {Chatterjee}},\ and\ \bibinfo {author} {\bibfnamefont {T.}~\bibnamefont
  {Speck}},\ }\bibfield  {title} {\enquote {\bibinfo {title} {Aggregation and
  sedimentation of active {Brownian} particles at constant affinity},}\ }\href
  {https://doi.org/10.1063/1.5081115} {\bibfield  {journal} {\bibinfo
  {journal} {J. Chem. Phys.}\ }\textbf {\bibinfo {volume} {150}},\ \bibinfo
  {pages} {064910} (\bibinfo {year} {2019})}\BibitemShut {NoStop}%
\bibitem [{\citenamefont {Pietzonka}\ \emph {et~al.}(2019)\citenamefont
  {Pietzonka}, \citenamefont {Fodor}, \citenamefont {Lohrmann}, \citenamefont
  {Cates},\ and\ \citenamefont {Seifert}}]{Pietzonka/etal:2019}%
  \BibitemOpen
  \bibfield  {author} {\bibinfo {author} {\bibfnamefont {P.}~\bibnamefont
  {Pietzonka}}, \bibinfo {author} {\bibfnamefont {E.}~\bibnamefont {Fodor}},
  \bibinfo {author} {\bibfnamefont {C.}~\bibnamefont {Lohrmann}}, \bibinfo
  {author} {\bibfnamefont {M.~E.}\ \bibnamefont {Cates}},\ and\ \bibinfo
  {author} {\bibfnamefont {U.}~\bibnamefont {Seifert}},\ }\bibfield  {title}
  {\enquote {\bibinfo {title} {Autonomous engines driven by active matter:
  Energetics and design principles},}\ }\href
  {https://doi.org/10.1103/PhysRevX.9.041032} {\bibfield  {journal} {\bibinfo
  {journal} {Phys. Rev. X}\ }\textbf {\bibinfo {volume} {9}},\ \bibinfo {pages}
  {041032} (\bibinfo {year} {2019})}\BibitemShut {NoStop}%
\bibitem [{\citenamefont {Speck}(2022)}]{Speck:2022}%
  \BibitemOpen
  \bibfield  {author} {\bibinfo {author} {\bibfnamefont {T.}~\bibnamefont
  {Speck}},\ }\bibfield  {title} {\enquote {\bibinfo {title} {Efficiency of
  isothermal active matter engines: Strong driving beats weak driving},}\
  }\href {https://doi.org/10.1103/PhysRevE.105.L012601} {\bibfield  {journal}
  {\bibinfo  {journal} {Phys. Rev. E}\ }\textbf {\bibinfo {volume} {105}},\
  \bibinfo {pages} {L012601} (\bibinfo {year} {2022})}\BibitemShut {NoStop}%
\bibitem [{\citenamefont {Gaspard}\ and\ \citenamefont
  {Kapral}(2017)}]{Gaspard/Kapral:2017JCP}%
  \BibitemOpen
  \bibfield  {author} {\bibinfo {author} {\bibfnamefont {P.}~\bibnamefont
  {Gaspard}}\ and\ \bibinfo {author} {\bibfnamefont {R.}~\bibnamefont
  {Kapral}},\ }\bibfield  {title} {\enquote {\bibinfo {title} {Communication:
  Mechanochemical fluctuation theorem and thermodynamics of self-phoretic
  motors},}\ }\href {https://doi.org/10.1063/1.5008562} {\bibfield  {journal}
  {\bibinfo  {journal} {J. Chem. Phys.}\ }\textbf {\bibinfo {volume} {147}},\
  \bibinfo {pages} {211101} (\bibinfo {year} {2017})}\BibitemShut {NoStop}%
\bibitem [{\citenamefont {Gaspard}\ and\ \citenamefont
  {Kapral}(2018)}]{gaspar:2018}%
  \BibitemOpen
  \bibfield  {author} {\bibinfo {author} {\bibfnamefont {P.}~\bibnamefont
  {Gaspard}}\ and\ \bibinfo {author} {\bibfnamefont {R.}~\bibnamefont
  {Kapral}},\ }\bibfield  {title} {\enquote {\bibinfo {title} {Fluctuating
  chemohydrodynamics and the stochastic motion of self-diffusiophoretic
  particles},}\ }\href {https://doi.org/10.1063/1.5020442} {\bibfield
  {journal} {\bibinfo  {journal} {J. Chem. Phys.}\ }\textbf {\bibinfo {volume}
  {148}},\ \bibinfo {pages} {134104} (\bibinfo {year} {2018})}\BibitemShut
  {NoStop}%
\bibitem [{\citenamefont {Huang}\ \emph {et~al.}(2018)\citenamefont {Huang},
  \citenamefont {Schofield}, \citenamefont {Gaspard},\ and\ \citenamefont
  {Kapral}}]{huang:2018}%
  \BibitemOpen
  \bibfield  {author} {\bibinfo {author} {\bibfnamefont {M.-J.}\ \bibnamefont
  {Huang}}, \bibinfo {author} {\bibfnamefont {J.}~\bibnamefont {Schofield}},
  \bibinfo {author} {\bibfnamefont {P.}~\bibnamefont {Gaspard}},\ and\ \bibinfo
  {author} {\bibfnamefont {R.}~\bibnamefont {Kapral}},\ }\bibfield  {title}
  {\enquote {\bibinfo {title} {Dynamics of {Janus} motors with microscopically
  reversible kinetics},}\ }\href {https://doi.org/10.1063/1.5029344} {\bibfield
   {journal} {\bibinfo  {journal} {J. Chem. Phys.}\ }\textbf {\bibinfo {volume}
  {149}},\ \bibinfo {pages} {024904} (\bibinfo {year} {2018})}\BibitemShut
  {NoStop}%
\bibitem [{\citenamefont {Gaspard}\ and\ \citenamefont
  {Kapral}(2019{\natexlab{a}})}]{Gaspard/Kapral:2019a}%
  \BibitemOpen
  \bibfield  {author} {\bibinfo {author} {\bibfnamefont {P.}~\bibnamefont
  {Gaspard}}\ and\ \bibinfo {author} {\bibfnamefont {R.}~\bibnamefont
  {Kapral}},\ }\bibfield  {title} {\enquote {\bibinfo {title} {The stochastic
  motion of self-thermophoretic {Janus} particles},}\ }\href
  {https://doi.org/10.1088/1742-5468/ab252f} {\bibfield  {journal} {\bibinfo
  {journal} {J. Stat. Mech.}\ }\textbf {\bibinfo {volume} {2019}},\ \bibinfo
  {pages} {074001} (\bibinfo {year} {2019}{\natexlab{a}})}\BibitemShut
  {NoStop}%
\bibitem [{\citenamefont {Gaspard}\ and\ \citenamefont
  {Kapral}(2019{\natexlab{b}})}]{Gaspard/Kapral:2019}%
  \BibitemOpen
  \bibfield  {author} {\bibinfo {author} {\bibfnamefont {P.}~\bibnamefont
  {Gaspard}}\ and\ \bibinfo {author} {\bibfnamefont {R.}~\bibnamefont
  {Kapral}},\ }\bibfield  {title} {\enquote {\bibinfo {title} {Thermodynamics
  and statistical mechanics of chemically powered synthetic nanomotors},}\
  }\href {https://doi.org/10.1080/23746149.2019.1602480} {\bibfield  {journal}
  {\bibinfo  {journal} {Adv. Phys. X}\ }\textbf {\bibinfo {volume} {4}},\
  \bibinfo {pages} {1602480} (\bibinfo {year}
  {2019}{\natexlab{b}})}\BibitemShut {NoStop}%
\bibitem [{\citenamefont {Gaspard}\ and\ \citenamefont
  {Kapral}(2020)}]{Gaspard/Kapral:2020}%
  \BibitemOpen
  \bibfield  {author} {\bibinfo {author} {\bibfnamefont {P.}~\bibnamefont
  {Gaspard}}\ and\ \bibinfo {author} {\bibfnamefont {R.}~\bibnamefont
  {Kapral}},\ }\bibfield  {title} {\enquote {\bibinfo {title} {Active matter,
  microreversibility, and thermodynamics},}\ }\href
  {https://doi.org/10.34133/2020/9739231} {\bibfield  {journal} {\bibinfo
  {journal} {Research}\ }\textbf {\bibinfo {volume} {2020}},\ \bibinfo {pages}
  {9739231} (\bibinfo {year} {2020})}\BibitemShut {NoStop}%
\bibitem [{\citenamefont {De~Corato}\ and\ \citenamefont
  {Pagonabarraga}(2022)}]{DeCorato/Pagonabarraga:2022}%
  \BibitemOpen
  \bibfield  {author} {\bibinfo {author} {\bibfnamefont {M.}~\bibnamefont
  {De~Corato}}\ and\ \bibinfo {author} {\bibfnamefont {I.}~\bibnamefont
  {Pagonabarraga}},\ }\bibfield  {title} {\enquote {\bibinfo {title} {Onsager
  reciprocal relations and chemo-mechanical coupling for chemically active
  colloids},}\ }\href {https://doi.org/10.1063/5.0098425} {\bibfield  {journal}
  {\bibinfo  {journal} {J. Chem. Phys.}\ }\textbf {\bibinfo {volume} {157}},\
  \bibinfo {pages} {084901} (\bibinfo {year} {2022})}\BibitemShut {NoStop}%
\bibitem [{\citenamefont {De~Groot}\ and\ \citenamefont
  {Mazur}(2013)}]{DeGroot/Mazur:2013}%
  \BibitemOpen
  \bibfield  {author} {\bibinfo {author} {\bibfnamefont {S.}~\bibnamefont
  {De~Groot}}\ and\ \bibinfo {author} {\bibfnamefont {P.}~\bibnamefont
  {Mazur}},\ }\href@noop {} {\emph {\bibinfo {title} {Non-Equilibrium
  Thermodynamics}}},\ Dover Books on Physics\ (\bibinfo  {publisher} {Dover
  Publications},\ \bibinfo {year} {2013})\BibitemShut {NoStop}%
\bibitem [{\citenamefont {Howse}\ \emph {et~al.}(2007)\citenamefont {Howse},
  \citenamefont {Jones}, \citenamefont {Ryan}, \citenamefont {Gough},
  \citenamefont {Vafabakhsh},\ and\ \citenamefont
  {Golestanian}}]{Howse/etal:2007}%
  \BibitemOpen
  \bibfield  {author} {\bibinfo {author} {\bibfnamefont {J.~R.}\ \bibnamefont
  {Howse}}, \bibinfo {author} {\bibfnamefont {R.~A.~L.}\ \bibnamefont {Jones}},
  \bibinfo {author} {\bibfnamefont {A.~J.}\ \bibnamefont {Ryan}}, \bibinfo
  {author} {\bibfnamefont {T.}~\bibnamefont {Gough}}, \bibinfo {author}
  {\bibfnamefont {R.}~\bibnamefont {Vafabakhsh}},\ and\ \bibinfo {author}
  {\bibfnamefont {R.}~\bibnamefont {Golestanian}},\ }\bibfield  {title}
  {\enquote {\bibinfo {title} {Self-motile colloidal particles: From directed
  propulsion to random walk},}\ }\href
  {https://doi.org/10.1103/PhysRevLett.99.048102} {\bibfield  {journal}
  {\bibinfo  {journal} {Phys. Rev. Lett.}\ }\textbf {\bibinfo {volume} {99}},\
  \bibinfo {pages} {048102} (\bibinfo {year} {2007})}\BibitemShut {NoStop}%
\bibitem [{\citenamefont {Dunderdale}\ \emph {et~al.}(2012)\citenamefont
  {Dunderdale}, \citenamefont {Ebbens}, \citenamefont {Fairclough},\ and\
  \citenamefont {Howse}}]{Dunderdale/etal:2012}%
  \BibitemOpen
  \bibfield  {author} {\bibinfo {author} {\bibfnamefont {G.}~\bibnamefont
  {Dunderdale}}, \bibinfo {author} {\bibfnamefont {S.}~\bibnamefont {Ebbens}},
  \bibinfo {author} {\bibfnamefont {P.}~\bibnamefont {Fairclough}},\ and\
  \bibinfo {author} {\bibfnamefont {J.}~\bibnamefont {Howse}},\ }\bibfield
  {title} {\enquote {\bibinfo {title} {Importance of particle tracking and
  calculating the mean-squared displacement in distinguishing nanopropulsion
  from other processes},}\ }\href {https://doi.org/10.1021/la301370y}
  {\bibfield  {journal} {\bibinfo  {journal} {Langmuir}\ }\textbf {\bibinfo
  {volume} {28}},\ \bibinfo {pages} {10997--11006} (\bibinfo {year}
  {2012})}\BibitemShut {NoStop}%
\bibitem [{\citenamefont {Pati\~{n}o}\ \emph {et~al.}(2018)\citenamefont
  {Pati\~{n}o}, \citenamefont {Arqu\'{e}}, \citenamefont {Mestre},
  \citenamefont {Palacios},\ and\ \citenamefont
  {S\'{a}nchez}}]{Patino/etal:2018}%
  \BibitemOpen
  \bibfield  {author} {\bibinfo {author} {\bibfnamefont {T.}~\bibnamefont
  {Pati\~{n}o}}, \bibinfo {author} {\bibfnamefont {X.}~\bibnamefont
  {Arqu\'{e}}}, \bibinfo {author} {\bibfnamefont {R.}~\bibnamefont {Mestre}},
  \bibinfo {author} {\bibfnamefont {L.}~\bibnamefont {Palacios}},\ and\
  \bibinfo {author} {\bibfnamefont {S.}~\bibnamefont {S\'{a}nchez}},\
  }\bibfield  {title} {\enquote {\bibinfo {title} {Fundamental aspects of
  enzyme-powered micro- and nanoswimmers},}\ }\href
  {https://doi.org/10.1021/acs.accounts.8b00288} {\bibfield  {journal}
  {\bibinfo  {journal} {Acc. Chem. Res.}\ }\textbf {\bibinfo {volume} {51}},\
  \bibinfo {pages} {2662--2671} (\bibinfo {year} {2018})}\BibitemShut {NoStop}%
\bibitem [{\citenamefont {Krapf}\ \emph {et~al.}(2018)\citenamefont {Krapf},
  \citenamefont {Marinari}, \citenamefont {Metzler}, \citenamefont {Oshanin},
  \citenamefont {Xu},\ and\ \citenamefont {Squarcini}}]{Krapf/etal:NJP2018}%
  \BibitemOpen
  \bibfield  {author} {\bibinfo {author} {\bibfnamefont {D.}~\bibnamefont
  {Krapf}}, \bibinfo {author} {\bibfnamefont {E.}~\bibnamefont {Marinari}},
  \bibinfo {author} {\bibfnamefont {R.}~\bibnamefont {Metzler}}, \bibinfo
  {author} {\bibfnamefont {G.}~\bibnamefont {Oshanin}}, \bibinfo {author}
  {\bibfnamefont {X.}~\bibnamefont {Xu}},\ and\ \bibinfo {author}
  {\bibfnamefont {A.}~\bibnamefont {Squarcini}},\ }\bibfield  {title} {\enquote
  {\bibinfo {title} {Power spectral density of a single {Brownian} trajectory:
  what one can and cannot learn from it},}\ }\href
  {https://doi.org/10.1088/1367-2630/aaa67c} {\bibfield  {journal} {\bibinfo
  {journal} {New J. Phys.}\ }\textbf {\bibinfo {volume} {20}},\ \bibinfo
  {pages} {023029} (\bibinfo {year} {2018})}\BibitemShut {NoStop}%
\bibitem [{\citenamefont {Squarcini}, \citenamefont {Solon},\ and\
  \citenamefont {Oshanin}(2022)}]{Squarcini/etal:NJP2022}%
  \BibitemOpen
  \bibfield  {author} {\bibinfo {author} {\bibfnamefont {A.}~\bibnamefont
  {Squarcini}}, \bibinfo {author} {\bibfnamefont {A.}~\bibnamefont {Solon}},\
  and\ \bibinfo {author} {\bibfnamefont {G.}~\bibnamefont {Oshanin}},\
  }\bibfield  {title} {\enquote {\bibinfo {title} {Spectral density of
  individual trajectories of an active {Brownian} particle},}\ }\href
  {https://doi.org/10.1088/1367-2630/ac44e6} {\bibfield  {journal} {\bibinfo
  {journal} {New J. Phys.}\ }\textbf {\bibinfo {volume} {24}},\ \bibinfo
  {pages} {013018} (\bibinfo {year} {2022})}\BibitemShut {NoStop}%
\end{thebibliography}
%
\end{document}